\documentclass[twocolumn]{aastex61}
\pdfoutput=1 
\usepackage{amsmath,amstext}
\usepackage[T1]{fontenc}
\usepackage{apjfonts} 
\usepackage[figure,figure*]{hypcap}
\usepackage{tabularx}

\usepackage{CJKutf8}

\shorttitle{DARKNESS}
\shortauthors{Meeker et al.}

\begin{document}

\title{DARKNESS: a Microwave Kinetic Inductance Detector integral field spectrograph for high-contrast astronomy}

\author{Seth R. Meeker}
\affiliation{Department of Physics, University of California Santa Barbara, Santa Barbara, California 93106, USA}
\affiliation{Jet Propulsion Laboratory, California Institute of Technology, 4800 Oak Grove Dr., Pasadena, California 91109, USA}
\author{Benjamin A. Mazin}
\affiliation{Department of Physics, University of California Santa Barbara, Santa Barbara, California 93106, USA}
\author{Alex B. Walter}
\affiliation{Department of Physics, University of California Santa Barbara, Santa Barbara, California 93106, USA}
\author{Paschal Strader}
\affiliation{Dominican School of Philosophy and Theology, Berkeley, California 94708, USA}
\author{Neelay Fruitwala}
\affiliation{Department of Physics, University of California Santa Barbara, Santa Barbara, California 93106, USA}
\author{Clint Bockstiegel}
\affiliation{Department of Physics, University of California Santa Barbara, Santa Barbara, California 93106, USA}
\author{Paul Szypryt}
\affiliation{Department of Physics, University of California Santa Barbara, Santa Barbara, California 93106, USA}
\author{Gerhard Ulbricht}
\affiliation{School of Cosmic Physics, Dublin Institute of Advanced Studies, Dublin 4, Ireland}
\author{Gr\'{e}goire Coiffard}
\affiliation{Department of Physics, University of California Santa Barbara, Santa Barbara, California 93106, USA}
\author{Bruce Bumble}
\affiliation{Jet Propulsion Laboratory, California Institute of Technology, 4800 Oak Grove Dr., Pasadena, California 91109, USA}
\author{Gustavo Cancelo}
\affiliation{Fermi National Accelerator Laboratory, Pine Street, Bataravia, Illinois 60510, USA}
\author{Ted Zmuda}
\affiliation{Fermi National Accelerator Laboratory, Pine Street, Bataravia, Illinois 60510, USA}
\author{Ken Treptow}
\affiliation{Fermi National Accelerator Laboratory, Pine Street, Bataravia, Illinois 60510, USA}
\author{Neal Wilcer}
\affiliation{Fermi National Accelerator Laboratory, Pine Street, Bataravia, Illinois 60510, USA}
\author{Giulia Collura}
\affiliation{Department of Physics, University of California Santa Barbara, Santa Barbara, California 93106, USA}
\author{Rupert Dodkins}
\affiliation{Department of Physics, University of Oxford, Oxford OX1 3RH, UK}
\author{Isabel Lipartito}
\affiliation{Department of Physics, University of California Santa Barbara, Santa Barbara, California 93106, USA}
\author{Nicholas Zobrist}
\affiliation{Department of Physics, University of California Santa Barbara, Santa Barbara, California 93106, USA}
\author{Michael Bottom}
\affiliation{Jet Propulsion Laboratory, California Institute of Technology, 4800 Oak Grove Dr., Pasadena, California 91109, USA}
\author{J. Chris Shelton}
\affiliation{Jet Propulsion Laboratory, California Institute of Technology, 4800 Oak Grove Dr., Pasadena, California 91109, USA}
\author{Dimitri Mawet}
\affiliation{Department of Astronomy, California Institute of Technology, 1200 E. California Blvd., Pasadena, California, 91125, USA}
\author{Julian C. van Eyken}
\affiliation{IPAC, Mail Code 100-22, California Institute of Technology, 1200 E. California Blvd., Pasadena, California, 91125, USA}
\author{Gautam Vasisht}
\affiliation{Jet Propulsion Laboratory, California Institute of Technology, 4800 Oak Grove Dr., Pasadena, California 91109, USA}
\author{Eugene Serabyn}
\affiliation{Jet Propulsion Laboratory, California Institute of Technology, 4800 Oak Grove Dr., Pasadena, California 91109, USA}

\begin{abstract}
We present DARKNESS (the DARK-speckle Near-infrared Energy-resolving Superconducting Spectrophotometer), the first of several planned integral field spectrographs to use optical/near-infrared Microwave Kinetic Inductance Detectors (MKIDs) for high-contrast imaging. The photon counting and simultaneous low-resolution spectroscopy
provided by MKIDs will enable real-time speckle control techniques and post-processing speckle suppression at framerates capable of resolving the atmospheric speckles that currently limit high-contrast imaging from the ground. DARKNESS is now operational behind the PALM-3000 extreme adaptive optics system and the Stellar Double Coronagraph at Palomar Observatory. Here we describe the motivation, design, and characterization of the instrument, early on-sky results, and future prospects.

\end{abstract}


\section{Introduction \& Motivation}
\label{sec:intro}
The field of exoplanet imaging is young and advancing rapidly. Early ground-based surveys doubled as testbeds for high-contrast instrumentation and algorithm development while putting strong constraints on the frequency of giant exoplanets at large separations ($\sim$10s to 100s of AU) from their host stars  (see \citet{Bowler2016} for a comprehensive review of these first-generation surveys). More recently, a new class of dedicated near-IR high-contrast instruments have come online, including Project 1640 at Palomar \citep[P1640;][]{Hinkley2011}, The Gemini Planet Imager at Gemini South \citep[GPI;][]{Macintosh2006, Macintosh2014}, the Spectro-Polarimetric High-contrast Exoplanet REsearch instrument at VLT \citep[SPHERE;][]{Dohlen2006, Zurlo2014}, and the Subaru Coronagraphic Extreme Adaptive Optics system at Subaru \citep[SCExAO][]{Jovanovic2015}, with the goal of imaging exoplanetary systems at separations below 10~AU.\footnote{e.g. GPI's inner working angle (IWA) in H-band is $\sim$1~AU for a star at 10~pc, with the deepest contrasts achieved by $\sim$5 or 6~AU.} However, these instruments continue to find fewer giant exoplanets than predicted \citep{Macintosh2006}, suggesting a possible discrepancy between the planet mass function extrapolated from radial velocity surveys and the true giant exoplanet mass function \citep{Bowler2016}. To fully understand the mass distribution of giant exoplanets and their formation mechanisms we must achieve deeper contrasts at smaller angular separations.

These specialized instruments have similar anatomies. First, an extreme Adaptive Optics (XAO) system corrects the atmosphere's distortion of the incoming light.\footnote{Here ``extreme" means thousands of actuators performing the wavefront correction, compared to only hundreds in the AO systems that early surveys used} After XAO the on-axis star light is removed with a coronagraph, leaving any off-axis light from a planet or disk mostly unaffected. Finally, the remaining light is sent to the science instrument backend for imaging, spectroscopy, and/or polarimetry depending on the particular system's science instrument suite. While the exact capabilities vary from one system to another, each of the systems listed above features an integral field spectrograph (IFS) among their science instruments that returns a spectrum at each spatial element in the final image \citep{Hinkley2008,Larkin2014,Claudi2008,Groff2014}. With this configuration the majority of the overwhelming starlight can be suppressed, revealing nearby faint companions or disk systems with simultaneous imaging and spectroscopy \citep{Oppenheimer2013}.

Unfortunately, AO correction is never perfect --- residual atmospheric aberrations and non-common path aberrations (NCPAs) between the wavefront sensor (WFS) and science arm result in some starlight escaping coronagraphic rejection, creating a pattern of coherent \emph{speckles} in the final image that resemble faint companions. Quasi-static speckles --- resulting from static aberrations such as from the instrument optics --- have long decorrelation times on the order of minutes to hours \citep{Hinkley2007}. Several mitigation strategies exist for these static/quasi-static aberrations. NCPAs can be partially calibrated out before going on-sky with phase retrieval or interferometric calibration \citep{Burruss2010, Cady2013}, and long-lived speckles are typically subtracted in post-processing using a variety of differential imaging techniques like angular differential imaging \citep[ADI;][]{Marois2006} or spectral differential imaging \citep[SDI or spectral deconvolution; ][]{Sparks2002,Crepp2011}. These differential imaging strategies utilize some form of image diversity to generate a reference point spread function (PSF) as an estimate for the quasi-static speckle pattern. However, speckles resulting from residual atmospheric aberrations have decorrelation times on the order of milliseconds to seconds \citep{Macintosh2005}, too fast to be resolved with conventional IFS exposures, and must simply be averaged into a smooth halo during long integrations.

The speckle noise that remains after post-processing imposes the state-of-the-art planet-star contrast limits across all ground-based high-contrast instruments. Overcoming this speckle barrier requires a wavefront correction scheme that employs a focal plane wavefront sensor (FPWFS) to capture NCPAs, operating at kHz frame rates to track even the atmospheric speckles \citep{Guyon2005}. This application necessitates the development of fast, low-noise, near-infrared focal plane detectors. Here we introduce DARKNESS, a critical testbed for one such technology: Microwave Kinetic Inductance Detectors (MKIDs). With the unique capabilities of MKIDs, DARKNESS (and its successors) can simultaneously serve as the FPWFS and low-resolution IFS for science data.

\subsection{Microwave Kinetic Inductance Detectors}
\label{sec:mkids}
MKIDs \citep{Day2003} are an emerging low temperature detector (LTD) technology that, through microwave multiplexing techniques and inherently simple geometric design, enables relatively low cost kilopixel, and potentially megapixel, arrays. Originally conceived for sub-millimeter astronomy applications, MKIDs have recently been developed for UV, optical, and near-IR (UVOIR) wavelengths \citep{Mazin2012,Marsden2012, Szypryt2017}. At these wavelengths MKIDs detect individual photons with time resolution of a few microseconds, are capable of measuring individual photon energies to within a few percent, and have no analogue for the read-noise or dark current present in conventional semiconductor-based detectors.

MKIDs operate on the principle that incident photons will briefly change the surface impedance of a superconductor through the kinetic inductance effect, caused by the finite mass of the Cooper pairs that compose the supercurrent. To reverse the supercurrent direction in an AC field requires the extraction of stored Cooper pair kinetic energy, which ultimately manifests as an additional surface inductance that dominates the overall surface impedance. When an incident photon breaks Cooper pairs the superconductor's surface impedance changes, and this change can be measured by using the superconductor as a variable inductor in a resonant circuit. The change in surface impedance will shift the frequency of the resonance, which, in turn, causes a shift in the amplitude and phase of a microwave probe signal that is coupled to the circuit. By fabricating individual resonators with different resonant frequencies a single microwave transmission line can read out thousands of resonators simultaneously through frequency domain multiplexing \citep{McHugh2012}.

Since the gap energy of a superconductor is very small (e.g. $\sim$10$^{-4}$~eV for Aluminum), a high energy ($\sim$1~eV) optical photon will break many thousands of Cooper pairs, creating a ``pulse" in the probe signal's phase as the resonator rapidly moves off-resonance (in $\sim$1~$\mu$s) upon absorbing the photon, then decays back to its steady state more slowly (in $\sim$20~$\mu$s) according to the quasiparticle lifetime. The sharpness of the pulse rise-time, combined with the continuous readout of every resonator simultaneously, provides the high time resolution of MKIDs. Photons of different energies will break different numbers of Cooper pairs, creating pulses in phase with height proportional to photon energy, which provides the intrinsic energy resolution of MKIDs.

The capabilities of UVOIR MKIDs have been demonstrated on-sky with the ARray Camera for Optical to Near-infrared Spectrophotometry \citep[ARCONS; ][]{Mazin2013}, a seeing-limited IFS designed for the Coud\`e focus at Palomar and Lick Observatories, featuring a 2024 pixel MKID array optimized for a 0.4~$\mu$m to 1.1~$\mu$m bandwidth. ARCONS was the first MKID camera at any wavelength to produce published astronomical science results \citep{Strader2013, Szypryt2014, Strader2016, Collura2017}. DARKNESS inherits significantly from its predecessor, ARCONS, especially in the MKID design (Section~\ref{sec:d3}), readout electronics (Section~\ref{sec:readout}), and analysis software (Section~\ref{sec:pipeline}).

\subsection{High-speed Speckle Correction Techniques}
\label{speckletechniques}

Here we mention a few speckle suppression techniques that will especially benefit from the advent of MKIDs for high-contrast applications. 

\vspace{-5pt}
\subsubsection{Statistical Speckle Discrimination}
\label{sec:ssd}
A significant body of work has been dedicated to the underlying probability density function (PDF) from which a speckle's intensities are drawn, which is known to be a modified Rician (MR) form given by
\begin{equation}
	\label{eq:mr}
p_{MR}(I) = \frac{1}{I_{s}}\text{exp}\bigg(-\frac{I+I_{c}}{I_{s}}\bigg)I_{0}\bigg(\frac{2\sqrt{I I_{c}}}{I_{s}}\bigg)
         \end{equation}
characterized by the static PSF contribution, \emph{I$_{C}$}, and random speckle intensities, \emph{I$_{S}$}, where $I_{0}(x)$ is the zero-order modified Bessel function of the first kind \citep{Perrin2003, Goodman2005, Aime2004, Fitzgerald2006, Soummer2007}. The ratio $I_{c}/I_{s}$ essentially characterizes the skewness, and the mean and variance of $I$ are:
\begin{equation}
\label{eq:mrmean}
\mu_{I} = I_{c}+I_{s}
\end{equation}

\begin{equation}
\label{eq:mrstd}
\sigma_{I}^{2} = I_{s}^{2} + 2I_{c}I_{s}
\end{equation}

When a large number of independent and identically distributed (i.i.d.) speckles are co-added, their statistics will become Gaussian by the central limit theorem, which is what most studies assume when quoting "5-$\sigma$" contrast curves. However, care must be taken when making this assumption, as improper treatment of the speckles' statistics (i.e. assuming Gaussianity when speckle noise is still correlated and thus the i.i.d. criterion is not true) can lead to severely underestimated false-alarm probablilities \citep{MaroisCL2008}.

Previously, studies have relied on whitening of the statistics, whereby PSF subtraction and other post-processing removes the correlated noise from (quasi-) static speckles. This leaves the atmospheric speckles to average together, and since they decorrelate quickly they can be considered i.i.d. and Gaussian statistics can be applied (see discussion in Section 1.1 of \citet{MawetCL2014}).  However, averaging of atmospheric speckles is exactly what we hope to avoid.

When imaging faster than the atmospheric speckle decorrelation time their MR statistics can be preserved, and this information can be used in post-processing to distinguish them from astrophysical sources. Methods like Dark-Speckle Imaging \citep[DS; ][]{Labeyrie1995, Boccaletti1998, Boccaletti2001} and Stochastic Speckle Discrimination \citep[SSD; ][]{Gladysz2008, Gladysz2009} rely on the fact that speckle intensities and companion intensities are drawn from different distribution functions (MR and Poissonian, respectively). By making a histogram of a pixel's intensity from thousands of successive short exposures a distribution containing only speckle contribution can be distinguished from one containing speckle+exoplanet. To implement these techniques optimally requires low-noise near-IR detectors capable of $\sim$ms exposures, a technology that was previously unavailable \citep{Boccaletti2001,Gladysz2008}, making them promising techniques to pursue with MKIDs.

\vspace{-5pt}
\subsubsection{Speckle Nulling}
\label{sec:nulling}

Speckle nulling \citep{Borde&Traub, Martinache2014, Bottom2016b} is one of several techniques for performing focal plane wavefront control, but is most attractive for our purposes because it lends itself most easily to a fast feedback loop. Since the pupil plane (where the DM resides) and the image plane are related by a Fourier Transform, a speckle’s position and intensity in the image plane gives the spatial frequency, angle, and amplitude of a corresponding sine wave on the DM. The only unknown is the speckle’s phase, which can be determined with probe patterns on the DM, stepping the expected waveform through just a few phase values and watching the target speckle dim or brighten. With the phase measured, all the necessary information is available to apply the inverse waveform on the DM that will destructively interfere with the speckle. This process can be implemented in a closed-loop to continuously probe and null a region of speckles to dig a “dark-hole” in the image. With a science camera capable of $\sim$kHz frame rates this technique can even cancel out atmospheric speckles in real time.

\section{DARKNESS Overview}
\label{sec:design}
DARKNESS is the first of several planned IFSs (see also MEC \citep{Meeker2015} and PICTURE-C \citep{Cook2015}) built to demonstrate the potential of MKIDs for high-contrast astronomy. Its baseline design is for operation at Palomar Observatory with the PALM-3000 \citep[P3K;][]{Dekany2013} XAO system and the Stellar Double Coronagraph \citep[SDC; ][]{Bottom2016a}. Here we provide an overview of the instrument as it currently operates with the SDC, while highlighting the flexible aspects of its design that enable its future travel to other observatories (see Section~\ref{sec:conclusions}). A summary of key instrument parameters is provided in Table~\ref{table:darkparams}.

\begin{figure}[t]
\begin{center}
\includegraphics[width=1.0\linewidth]{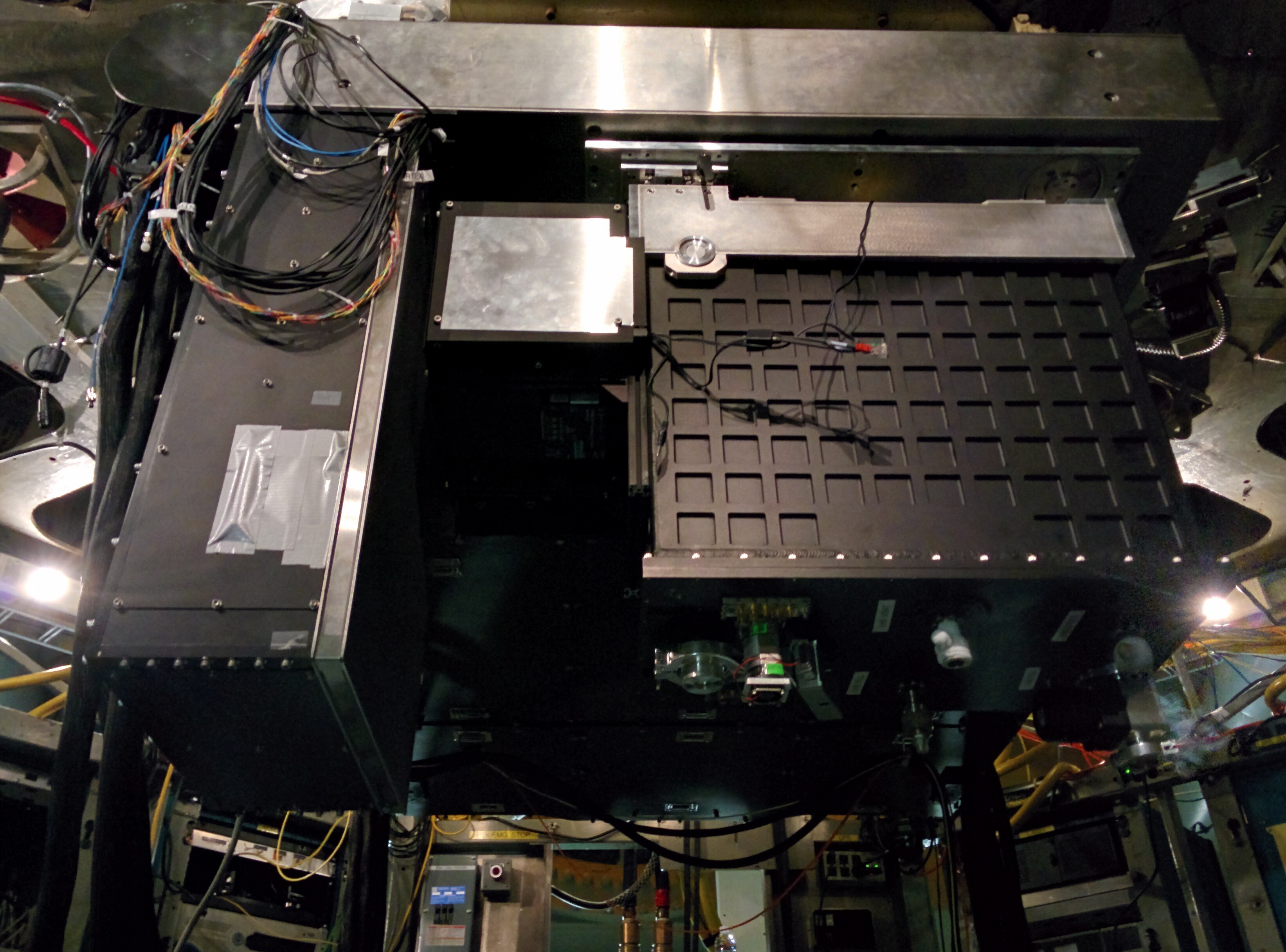}
\end{center}
\caption{DARKNESS (on the right) and the SDC (on the left) attached to the P3K bench and installed in the Hale Telescope Cassegrain cage. DARKNESS is shown here with its foreoptics box attached, but before connecting external cabling. The SDC electronics board typically hangs next to DARKNESS as well, but is removed here for clarity.} \label{fig:darkatcass}
\end{figure}

\begin{table}[t]
\centering
\caption{DARKNESS Overview}
\label{table:darkparams}
\begin{tabularx}{\linewidth}{@{}l *5{>{\centering\arraybackslash}X}@{}}
\hline
\hline
Parameters & Values\\ 
\hline
MKID Array Materials & PtSi on Sapphire substrate w/Nb ground plane\\
Array Format & 80$\times$125 pixels\\
Pixel Pitch & 150 $\mu$m\\
Plate Scale & 19.8 mas/pix\\
Wavelength Coverage & 0.8 to 1.4~$\mu$m\\
Spectral Resolution ($\lambda/\Delta\lambda$) & 7 to 5\\
Operating Temperature & 100 mK\\
Cryostat 100~mK Hold Time & 13 hours\\
Cryostat 4~K \& 77~K Hold Times & 40 hours\\
Cryostat Dimensions (L$\times$W$\times$H cm) & 66 $\times$ 33 $\times$ 60\\
Cryostat Weight & 110~kg\\
\hline
\end{tabularx}
\end{table}

\subsection{Cryostat}
\label{sec:cryostat}
DARKNESS's cryostat is a liquid cryogen pre-cooled Adiabatic Demagnetization Refrigerator (ADR) capable of reaching temperatures below 100 mK. The custom dewar, built by Precision Cryogenics, is manufactured mostly from 6061-T6 aluminum and is designed as a drop-in replacement for the SDC's usual backend imager, PHARO \citep{Hayward2001}. The ADR unit from High Precision Devices was integrated into the dewar at UCSB. The complete cryostat measures roughly 66~cm long $\times$ 33~cm wide $\times$ 60~cm tall, filling a similar envelope on the P3K bench as PHARO, but with a few extra inches of height to accommodate extra cryogen volume. The weight (when filled) is roughly 110~kg. A detailed accounting of the cryostat internals can be found in \citet{Meeker2015}. We briefly summarize key features and measured performance here.

A liquid cryogen design was selected due to space constraints in the Hale Telescope Cassegrain cage, preventing the use of pulse tube cooling which would require heavy and rigid drag-lines. Internal to the 300~K vacuum shell is an 8~liter LN2 tank that maintains a layer of thermal radiation shielding at 77~K for $\sim$40~hour hold time on a single fill. Nested inside the 77~K shield is a 24~liter LHe tank and another layer of thermal radiation shielding surrounding the 4~K experimental volume where the ADR unit and detector package reside. The 4~K stage has a similar $\sim$40~hour hold time per fill. 

The ADR acts as a single-shot magnetic cooler that brings the MKID array down to 100~mK where the temperature is stabilized with a feedback loop to the ADR magnet power supply. We achieve a 100~mK hold time of 13~hours on-sky, more than sufficient for a night of calibrations and observations. Special care has been taken to isolate and shield the MKID array from the ADR's magnetic field. The detector package and ADR are mounted to the 4~K stage far apart, with the 100~mK ADR cold finger attached to the MKID array via a flexible copper strap, and the detector package is enclosed in an Amumetal magnetic shield. 

\begin{figure*}[]
\begin{center}
\includegraphics[width=1.0\textwidth]{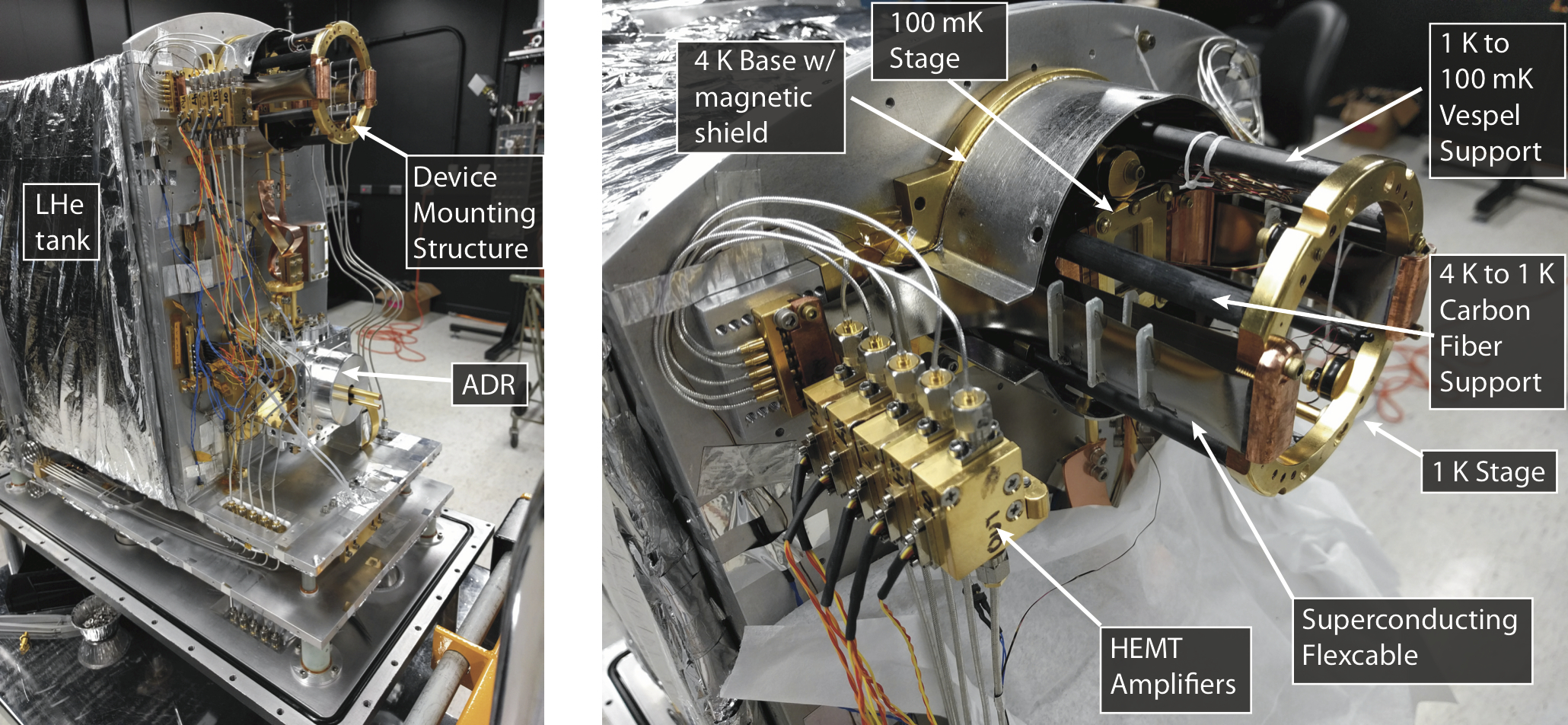}
\end{center}
\caption{(Left) Photograph of DARKNESS cryostat with 300 K vacuum shell and 77 K and 4 K radiation shields removed to show 4 K experiment volume with ADR and device mounting stage. (Right) Close-up of the device mounting structure with magnetic shield and 1 K baffle removed.} \label{fig:cryostat}
\end{figure*}

The detector package is comprised of three stages: a base plate at 4~K that also holds the magnetic shield, an intermediate 1~K ring that stands off from the 4~K base on Carbon fiber supports, and the 100~mK stage suspended from the 1~K ring by hollow Vespel SCP-5050 rods (Figure \ref{fig:cryostat} (Right)). This design thermally isolates the 100~mK stage, places the MKID array far from the magnetic shield opening where field leakage will be strongest, and also accommodates a 1~K baffle to block scattered light and 4~K blackbody radiation that could increase the phase-noise floor of the detector. 

\subsubsection{Mounting to the AO bench}
DARKNESS is held in a custom mounting cradle by three pins near the top of the vacuum shell: two on the sides near the front and one on the rear face. This cradle closely follows the design of the P1640 IFS mounting bracket \citep{Hinkley2011}, including $\pm$10~degrees of pitch adjustment using a screw-jack at the rear of the instrument and $\sim$1~inch of focus adjustment by hanging the instrument from three Bosch-Rexroth ball rails. These runners are then attached to an Aluminum 7071 mounting plate that screws directly to the P3K bench.

\subsubsection{Wiring}
\noindent\textbf{Microwave Signal Path:} DARKNESS requires five feedlines to read out an entire array.  The signal paths begin with hermetic SMA bulkhead connectors, bringing the signals in through the bottom face of the cryostat. Five laser welded stainless steel 0.087~inch semi-rigid coax cables bring the signal from the inside of the 300~K shell to the 4~K stage, with a heat sinking clamp heatsinking the outer conductor only at 77~K along the way. From the feedthroughs at the bottom of the 4~K stage the signals pass through 20 dB attenuators for reducing room temperature Johnson noise, then hand-formable SMA-to-SMA cables bring the signals to the MKID mounting structure. From here, SMA-to-G3PO cables connect to a superconducting 53 wt\%~Niobium/47 wt\%~Titanium (NbTi) flex cable.

We have fabricated custom 0.096~mm thick microstrip NbTi/polyimide/NbTi flex cables to allow for a high density of feedlines while minimizing heat load from 4~K to 100~mK as compared to five individual NbTi coax. An example of this flex cable is visible in Figure \ref{fig:cryostat} (Right), connecting from the 4~K base of the device mounting structure to the 100~mK stage with an intermediate heat sink at the 1~K stage. A full report of their design and performance is available in \citet{Walter2017}. 

These flex cables connect to the MKID box through five small G3PO connectors allowing for a much more compact detector package than standard SMA connectors. The box-mounted G3PO connectors are solder connected to gold-plated copper on duroid co-planar waveguide (CPW) transition boards, which are then aluminum wire-bonded to the MKID chip. After passing through the MKID array the five signals are brought out through the same series of CPW board, G3PO connectors, custom NbTi microstrip flexcable, and G3PO-to-SMA wires. At 4~K each line is amplified by a 4--8 GHz Low Noise Factory High Electron Mobility Transistor (HEMT) amplifier with a noise temperature of 2~K. Hand-formable SMA-to-SMA cables again bring the signal to the bottom of the 4~K plate, and stainless steel coax take the signals from there to 300~K.

\noindent\textbf{DC wiring:} DARKNESS has two 24-pin DC wire bundles going to 4~K to provide HEMT biasing and thermometry.  Current is supplied to the ADR magnet by two DC leads using copper wire from 300~K to 77~K, high-T$_{c}$ superconductor from 77~K to 4~K, and superconducting (NbTi) wire from 4~K to the magnet.

\subsection{MKID Array}
\label{sec:d3}

\begin{figure*}[]
\begin{center}
\includegraphics[width=\textwidth]{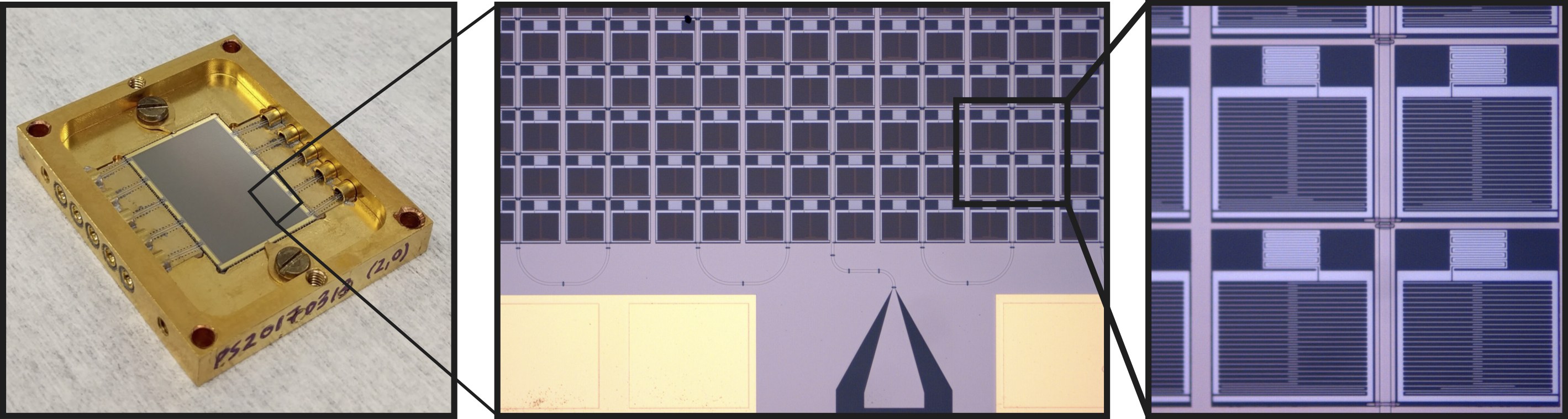}
\end{center}
\caption{(Left) A 10,000 MKID PtSi on Sapphire D-3 array mounted in its microwave package. (Center) Detail image of the array showing the CPW transmission line with bondpad for one feedline. (Right) Further detail of several MKID pixels. The densely meandered patches at the top of each pixel are the photosensitive inductors, and the large sparse sections are the interdigitated capacitors used to tune each MKID to a unique resonant frequency.} \label{fig:d3}
\end{figure*}

DARKNESS houses a 10,000 pixel (80x125) MKID array with 150~$\mu$m pixel pitch, optimized for a 0.8 $\mu$m to 1.4 $\mu$m bandwidth. DARKNESS pixels follow a similar lumped-element design to those used in ARCONS \citep{Marsden2012,Mazin2013}, and early array designs presented in \citet{Meeker2015} were made out of titanium nitride (TiN) on Silicon (Si) substrates and also followed a similar layout and fabrication process to ARCONS arrays. \citep{Leduc2010,Mazin2012}. The most recent DARKNESS arrays (those that produced the on-sky results presented in Section~\ref{sec:commissioning}) follow similar design principles, but take advantage of recent work exploring platinum silicide (PtSi) as an alternative to TiN \citep{Szypryt2016}. Figure~\ref{fig:d3} shows the PtSi on Sapphire array (known as D-3) used on DARKNESS's most recent observing run, mounted in its microwave package with accompanying microscope photos revealing details of the individual pixels. A full reporting of the design, fabrication, and performance of these PtSi arrays is presented in \citet{Szypryt2017}. While resonator internal quality factor, $Q_{i}$, is high in single-layer test chips --- comparable to the values regularly achieved with TiN ---  work is ongoing to preserve high quality through the full array fabrication process. Nonetheless, our current D-3 arrays demonstrate sensitivity and energy resolution on-par with or better than that seen in DARKNESS's early TiN arrays (see Section~\ref{sec:verification}), with dramatically improved fabrication uniformity \citep{Szypryt2016} and photometric stability (hot-pixel behavior in our TiN arrays is quantified in \citet{vanEyken2015}, and elimination of this behavior with PtSi will be investigated fully in a forthcoming manuscript).

\subsection{Optics}
\label{sec:optics}

\subsubsection{Cryostat Optics}
DARKNESS's cryostat optics are very simple, requiring only an entrance window, a pair of cold, IR-blocking filters at the 77~K and 4~K stages, and a microlens array (MLA) that concentrates the light from the final image plane onto the photo-sensitive inductor of each pixel. 

The MLA from Advanced Microoptic Systems is composed of roughly 100~x~145 lenslets with 150~$\mu$m pitch, and made from 1~mm thick STIH53 glass.

\begin{figure}[t]
\begin{center}
\includegraphics[width=1.0\linewidth]{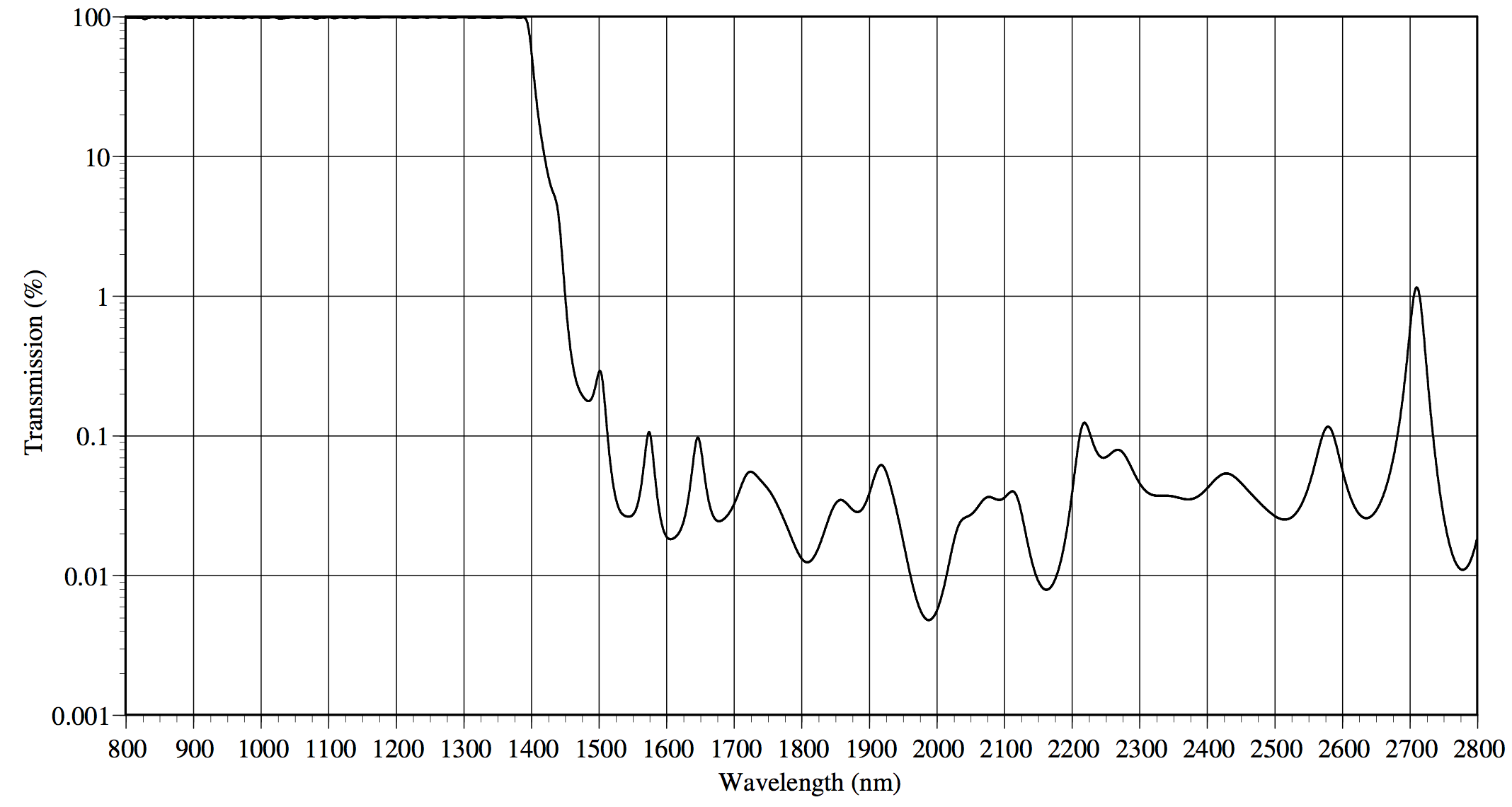}
\end{center}
\caption{Theoretical transmission curve of one IR-blocking/bandpass filter, courtesy of Custom Scientific.  The transmission of the thick N-BK7 substrate is not included. Blocking is specified as T$_{Avg}\leq$~0.4\% from 1.5 to 2.8~$\mu$m. Our filter stack combines two such filters, providing roughly OD4 blocking in the specified range.} \label{fig:bandpass}
\end{figure}

The entrance window is 12.5~mm thick AR-coated fused Silica. The 77~K and 4~K windows are both N-BK7 glass, 10~mm and 20~mm thick, respectively. These cold windows are coated with a custom IR-blocking filter from Custom Scientific, and define the cutoffs of our observing band. Since these are reflective coatings, both windows are mounted at 3$^{\circ}$ relative to the incident beam to reduce ghosting. The transmission curve of a single filter is shown in Figure~\ref{fig:bandpass}. A first iteration of these blocking filters was not as stringent, and we were sensitive to a small, but noticeable population of 300~K blackbody photons where the filter transmission increased slightly around 2.7 $\mu$m. With the final filter curves presented here, when combined with the IR blocking ability of the N-BK7 substrates above 2.8 $\mu$m, we expect $<$~1 photons per pixel per second of 300~K blackbody leak within DARKNESS’s bandpass and $<$~15 photons per pixel per second below 4 $\mu$m. The level of IR blocking provided by the filters obviates the need for a cold pupil stop.

\subsubsection{Relay Optics}
\begin{figure}[t]
\begin{center}
\includegraphics[width=1.0\linewidth]{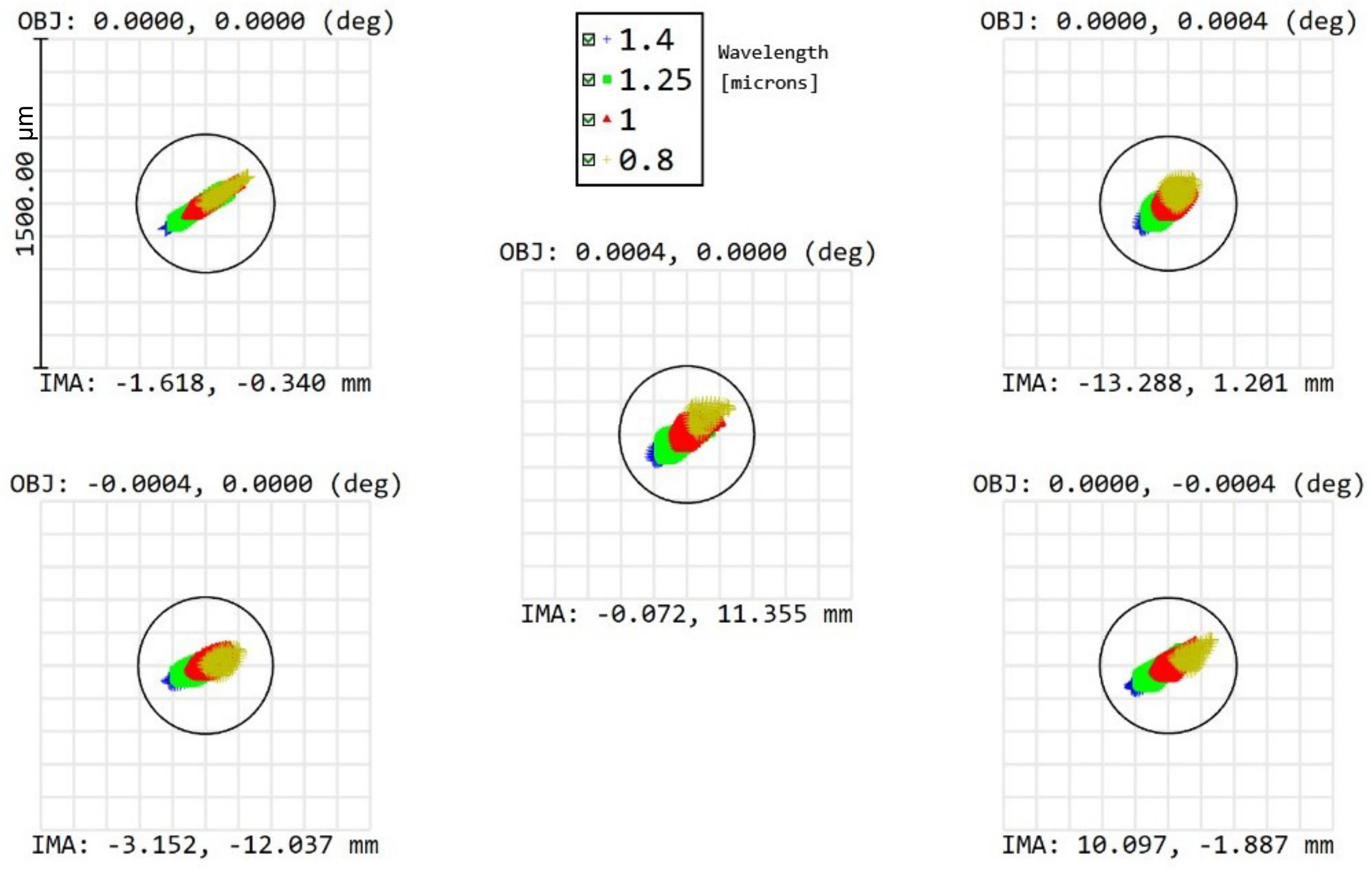}
\end{center}
\caption{Spot diagrams from several fields roughly covering the extent of the DARKNESS FOV. Airy radius is shown as a black circle, calculated at 0.8~$\mu$m. These diagrams show that even at the edges of the array the optical performance of the system will be diffraction limited (or rather, governed by the wavefront quality achieved with P3K) and chromatic effects are negligible. With an Airy radius of $\sim$300~$\mu$m at our shortest wavelength we expect Nyquist sampling (or better) of the diffraction limited PSF across our band with our 150~$\mu$m pixel pitch.} \label{fig:spots}
\end{figure}

The majority of DARKNESS's optical complexity is in the warm re-imaging optics that convert the SDC's \emph{f}/15.7 output beam to the \emph{f}/322 required for DARKNESS. The main constraints to this optical design are the diffraction limit of the telescope at DARKNESS's operating wavelengths and the need for telecentricity to ensure proper functioning of DARKNESS's microlens array. We performed Zemax simulations of the full optical train, including the Hale Telescope, P3K, the SDC, and DARKNESS, targeting Nyquist sampling of the diffraction beamwidth ($\lambda$/D) across our observing band. Resulting spot diagrams are shown in Figure~\ref{fig:spots}.

\begin{figure}[t]
\begin{center}
\includegraphics[width=1.0\linewidth]{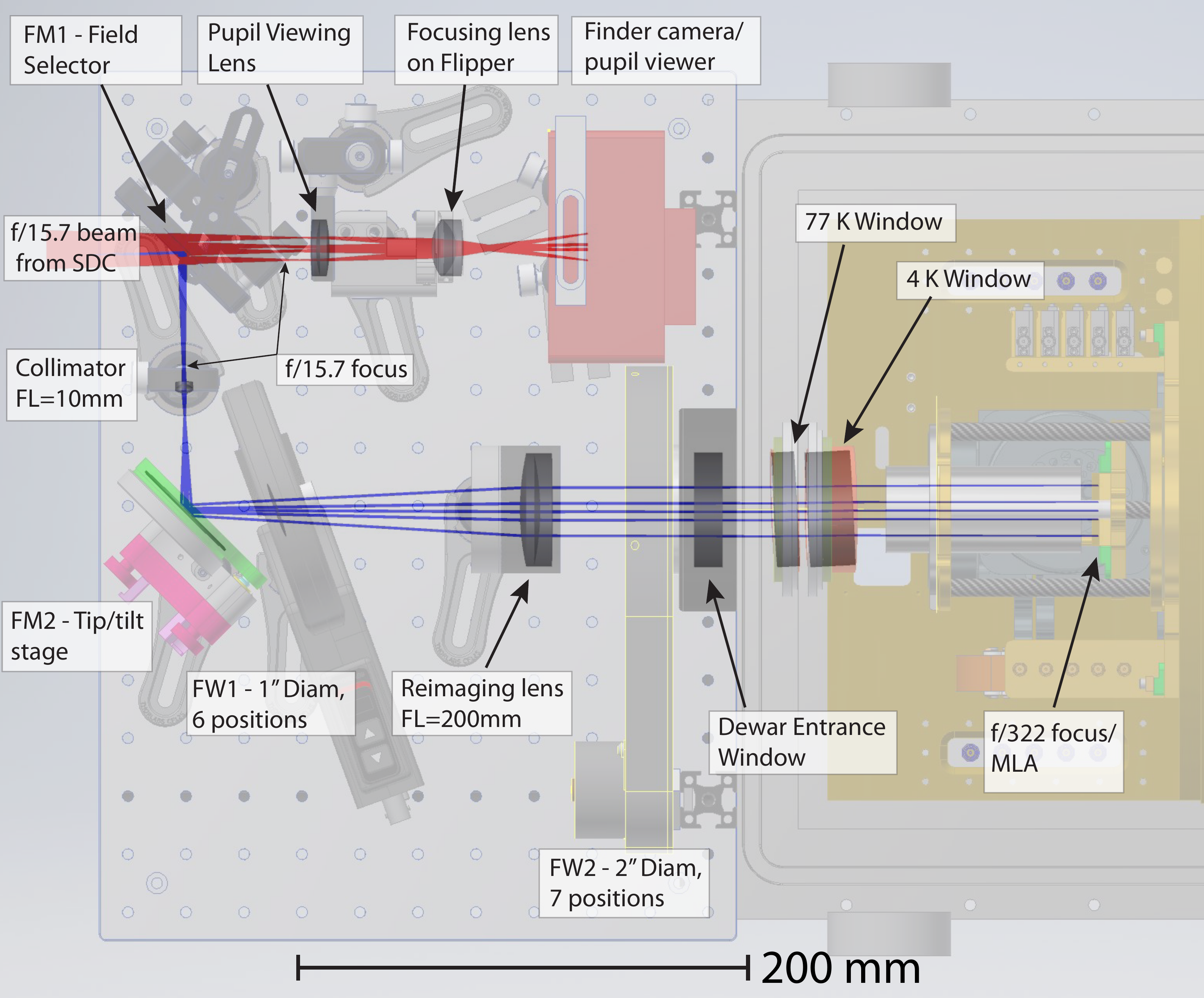}
\end{center}
\caption{DARKNESS foreoptics layout when operating with the SDC.} \label{fig:foreoptics}
\end{figure}

The layout of the fore-optics enclosure, including a ray trace of the re-imaging optics and a finder camera/pupil imaging arm, is shown in Figure~\ref{fig:foreoptics}. The \emph{f}/15.7 beam is first folded by a "field selector" comprised of a reflective aluminum rectangle deposited on the center of a BK7 window. This optic sends the central 3"x4" of the FOV to the \emph{f}/322 re-imaging optics and DARKNESS, while passing the surrounding full FOV to an SBIG STF-8300M CCD camera. The SBIG arm can switch between imaging (for target acquisition) and pupil viewing mode by flipping in an optional lens. The science beam is collimated and then folded again toward DARKNESS's entrance window. This fold mirror is on a remote controlled 3-axis Picomotor stage, allowing for fine adjustment of the FOV on the MKID array and automated dithering routines to fill in dead pixels. After the Picomotor mirror, the collimated beam passes through a 6-position 1" filter wheel with a selection of neutral density filters and is then re-focused to an \emph{f}/322 beam. A second, 7-position 2" filter wheel is placed just before the DARKNESS front window, providing a selection of color filters and also serving as the instrument's "shutter." A summary of the selectable filters can be found in Table~\ref{table:optics}.

This entire fore-optics enclosure is easily removable to accomodate a variety of observing configurations\footnote{For example: directly accepting the P3K beam, operating in a seeing-limited mode at the Cassegrain focus, or traveling to other observatories.} without disrupting the optical configuration used with SDC or making any changes to cold optics. Focus placement after removing/replacing the fore-optics is repeatable to within the day-to-day focus drift experienced over a several day observing run.

\subsubsection{Wavelength Calibration}
\label{sec:wvlcal}
As part of our calibration procedure, we must map phase offset to photon energy for each pixel using the measured response from a series of known laser sources. We employ a similar setup to the one described in \citet{Mazin2013} for ARCONS using a custom package that holds several laser diodes controlled remotely via Arduino. The lasers used for DARKNESS operate at 808~nm, 920~nm, 980~nm, 1120~nm, and 1310~nm. The diodes shine into an integrating sphere with fiber output, allowing us to mount the laser box assembly wherever is convenient (typically on the SDC electronics board alongside DARKNESS) while using a fiber to bring the light into our fore-optics box. 

In the fore-optics this fiber is installed next to the re-imaging lens, then directed at the DARKNESS front window to simply flood illuminate the detector with the help of a diffuser in FW2. Here uniformity is not a priority as much as decent count rate on every pixel. During observations, typically while tuning AO on a new target, we perform a wavelength calibration by closing FW1, moving FW2 to the diffuser position, then cycling through the lasers taking $\sim$1 minute of data from each. In processing, each laser peak is fit by a Gaussian to locate the mean phase offset from that wavelength for each pixel. The five peak locations are subsequently fit with a second-order polynomial to provide a complete mapping of phase-offset back to wavelength, which is applied as the first calibration step in our processing pipeline. This procedure is essentially unchanged from \citet{vanEyken2015}.

\subsubsection{Changes to the SDC}
\label{sec:sdc}

The SDC is a flexible coronagraph platform that features two internal focal planes and two pupil planes for deploying a variety of coronagraphic configurations, including a dual vector vortex coronagraph (VVC) designed to overcome diffraction from the secondary and spider obscurations in the pupil \citep{Bottom2016a, Mawet2005, mawet_improved_2011}. To optimally utilize a VVC requires superb correction from the XAO system, and very high Strehl ratio. Currently, P3K does not provide adequate Strehl ratio below J-band to justify the use of a VVC at these wavelengths (though this may change very soon - see discussion in Section~\ref{sec:conclusions}). 

\begin{figure}[]
\begin{center}
\includegraphics[width=1.0\linewidth]{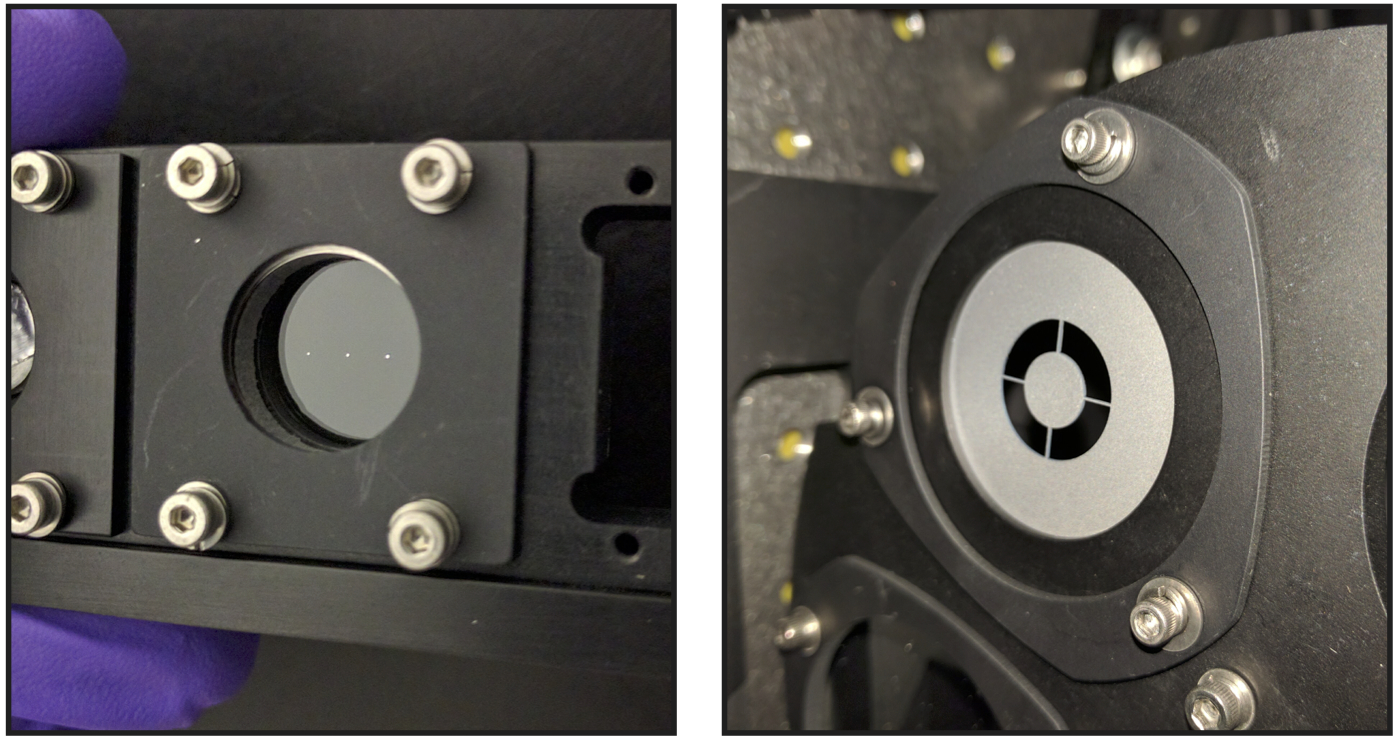}
\end{center}
\caption{Lyot Coronagraph optics in the SDC. (Left) Focal plane masks (Right) Lyot mask} \label{fig:lcmasks}
\end{figure}

With this consideration, and to minimize time-consuming coronagraph alignment while debugging DARKNESS, we fabricated and installed a simple Lyot coronagraph for use in SDC. The focal plane mask (FPM) is actually a set of three reflective aluminum spots of various diameters, sputter deposited on a single fused silica substrate, and located at the SDC's first focal plane. Similar to the standard SDC configuration, this FPM is installed on a linear translation stage along with a fiber ferrule, co-focused with the FPM, that is used to focus DARKNESS relative to the coronagraph focal plane. The FPM assembly attaches to the linear stage via magnetic mounts that are accurate to a few microns, allowing for easily reproducible alignment when swapping between our Lyot configuration and the SDC's standard VVCs \citep{Bottom2016a}. A Lyot mask, fabricated by deep reactive ion etching through a Silicon substrate, is installed at the SDC's first pupil plane. This mask can easily be made reflective to use the rejected starlight for low-order wavefront sensing (LOWFS) in future upgrades to the SDC. The secondary focal and pupil planes in the SDC are not used. With a selection of FPM diameters and Lyot mask sizes we can choose the desired configuration based on observing conditions, and this flexibility also allows for easy re-configuration when optimizing for different observing wavelengths. The masks used during commissioning, shown in Figure~\ref{fig:lcmasks} with selectable parameters listed in Table~\ref{table:optics}, were optimized for J-band operation since this is where we expected the best correction from P3K and focused the majority of commissioning observations.

The SDC includes an internal InGaAs quad-cell detector near its first focal plane that maintains very precise alignment of the target star on the coronagraph FPM. In SDC's standard configuration --- optimized for K-band observations --- a dichroic sends all J-band light to this IR tracker. When operating with DARKNESS we replace the dichroic with a reflective 1450~nm shortpass filter. A similar swap is also made in P3K where the typical 960~nm dichroic that splits WFS and science light would take part of DARKNESS's bandwidth at the blue end for the WFS. For DARKNESS observations we replace this with a 750~nm dichroic.

\begin{table}[]
\centering
\caption{Selectable optics parameters}
\label{table:optics}
\begin{tabularx}{\linewidth}{@{}l *5{>{\centering\arraybackslash}X}@{}}
\hline
\hline
Parameters & Values\\ 
\hline
\multicolumn{2}{c}{\textit{Lyot Coronagraph}}\\
FPM Diameters ($\lambda$/D at 1.25 $\mu$m) & 5, 6.6, 8.2\\
Undersized Lyot Stop Factor & 10\%, 15\%, 20\%\\

\multicolumn{2}{c}{\textit{Foreoptics}}\\
FW1 Neutral Density Filters & Closed, OD~3.0, OD~2.5, OD~1.5, OD~0.5, Open\\
FW2 Diffusers \& Color Filters & Closed, Y, z$_{s}$, Mauna~Kea~J, Diffuser~(high~grit), Diffuser~(low~grit), Open\\ 
\hline
\end{tabularx}
\end{table}

\subsection{Readout Electronics}
\label{sec:readout}

\begin{figure*}[t]
\begin{center}
\includegraphics[width=1.0\textwidth]{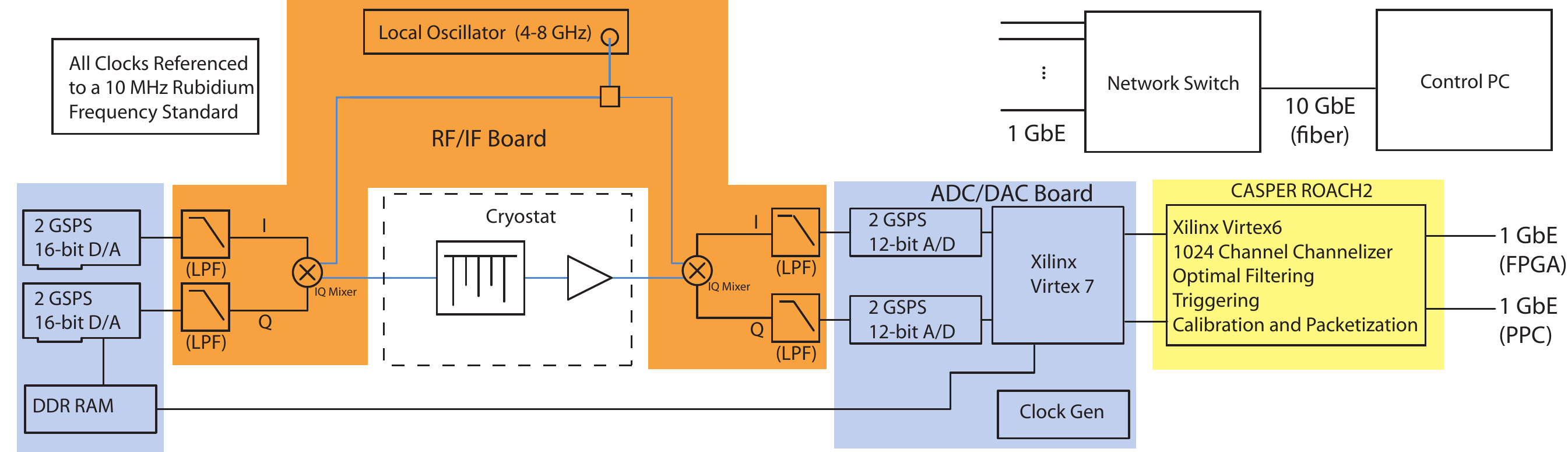}
\end{center}
\caption{Block diagram of the second generation UVOIR MKID readout. The blue, orange, and yellow shaded functions take place on the ADC/DAC, IF, and ROACH2 boards, respectively.} \label{fig:readoutbd}
\end{figure*}

DARKNESS's readout hardware and photon detection firmware are an evolution of those used in ARCONS, and follow the same strategy as that outlined in \citet{McHugh2012}. In general, these instruments use a heterodyne mixing scheme, where a set of probe tones is generated, one for each MKID resonator.  These tones are then passed through the device where the effects of illumination on the MKID array are imprinted on the probes, and this altered signal is compared against the original to detect the individual photon strikes. In the DARKNESS readout scheme, frequency comb generation/conversion and photon detection are handled with a combination of three boards: a ROACH2 board, a combination ADC/DAC board, and an intermediate frequency (IF) board. ROACH2 is the second generation of the CASPER Reconfigurable Open-Architecture Computing Hardware (ROACH), a platform originally intended for radio astronomy, and selected for our purposes for the flexibility and rapid development it enables.\footnote{More information on ROACH2 can be found on the CASPER wiki: casper.berkeley.edu/wiki/ROACH2} The additional resources provided by ROACH2 and the advances in ADC/DAC technology since ARCONS development has enabled a substantial increase in readout density. One set of DARKNESS boards is capable of reading out 1024 MKIDs in 2~GHz bandwidth, so each DARKNESS feedline requires two sets of boards for resonators covering 4~GHz of bandwidth, for a total of 10 sets for 10,000 pixels. For comparison ARCONS required eight ROACH boards for 2024 pixels. We summarize DARKNESS's board specifications, readout signal chain, and photon detection here, but encourage interested readers to consult \citet{StraderThesis} for significantly more detailed descriptions.

Figure \ref{fig:readoutbd} provides a block diagram of the readout chain. Definition of the tone frequencies and backend signal processing are handled on the Virtex 6 field programmable gate array (FPGA) on the ROACH2 board.  The probe tones for each resonator are created as complex waveforms at low frequency, generated as separate real (I or in-phase) and imaginary (Q or quadrature) components using dual 2~GSPS (giga-samples per second) 16-bit DACs. These I and Q components are then combined on the IF board and also mixed with a local oscillator (LO) up to our MKID frequencies (4~to~8.5~GHz). The summed waveforms (representing a "comb" in frequency space) are then sent to DARKNESS where they pass through the MKID, are amplified at 4~K with HEMT amplifiers, then brought back to the readout electronics. After another round of amplification on the IF board, the signals are mixed down to base-band and are broken back into I and Q components, then sent to the ADC/DAC board for digital conversion with dual 2~GSPS 12-bit ADCs. Finally this I and Q data is sent to the ROACH2 for channelization, filtering, and photon detection.

In the ROACH2 firmware, individual tones are separated out and downconverted to 0 Hz with a two stage channelization process. The I and Q data for each channel is low pass filtered, then converted to phase\footnote{$\phi = arctan(\frac{I-I_{center}}{Q-Q_{center}})$ where ($I_{center}$, $Q_{center}$) is the center of the resonator loop in the I/Q plane.} and filtered with a finite impulse response (FIR) filter with coefficients customized to each channel's unique pulse shape.  This shape is determined using the formalism of Optimal (Weiner) Filtering. After filtering, phase excursions that pass some threshold\footnote{Typically 4$\sigma$ from zero phase, where $\sigma$ is the standard deviation in optimally filtered phase noise for each pixel.} are flagged as photon events, and the photon is stored in a buffer as a 64-bit word that includes the arrival timestamp, phase offset of the event, and resonator ID for the pixel where the photon was detected. Every 0.5~ms or 100 photons (whichever comes first) the photons in every ROACH buffer are sent over 1 GbE connection to a HP Procurve switch.  The switch is connected to the data acquisition (DAQ) computer with a 10 GbE fiber link.  The computer collects the photons packets and writes them to a 80 TB RAID6 array continuously.  It also computes a quicklook image which it writes to disk every second.

\subsubsection{Electronics Rack}
\label{sec:rack}
\begin{figure}[t]
\begin{center}
\includegraphics[width=0.75\linewidth]{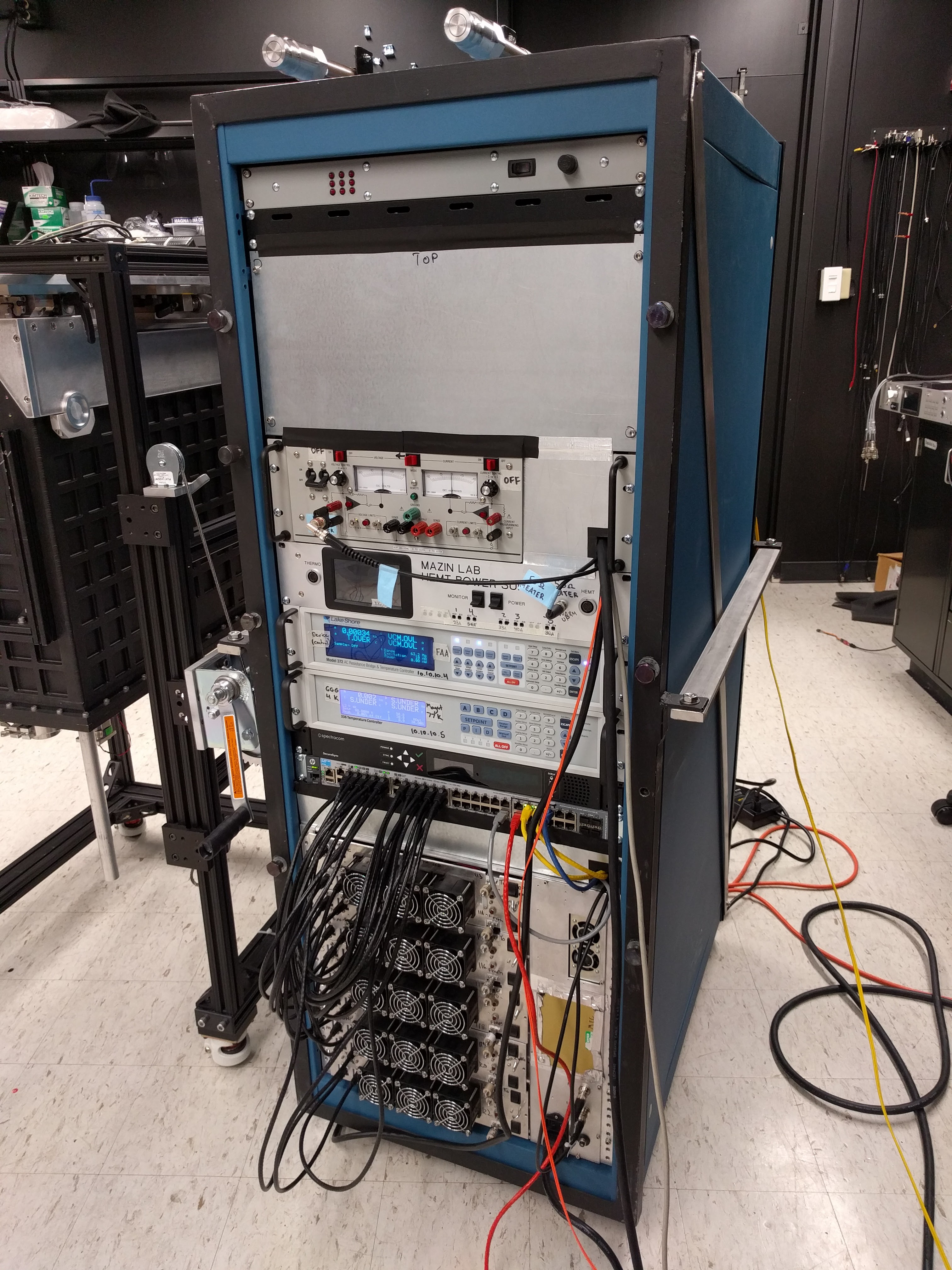}
\end{center}
\caption{DARKNESS electronics rack in lab at UCSB. This rack holds 5 readout cartridges with 10 sets of the readout boards described in Section~\ref{sec:readout}, 2 per cartridge, installed in a custom readout crate at the bottom of the rack that distributes power, 1 PPS, and 10 MHz signals to the boards. Above the crate is a network switch, time source, thermometry control, HEMT power supply, and ADR magnet power supply.} \label{fig:erack}
\end{figure}

DARKNESS's readout electronics, ADR magnet power supply, HEMT power supply, thermometry control, and GPS reference time source are installed in an electronics rack that attaches to the outside of the Cassegrain cage during observations (see Figure~\ref{fig:erack} for a view of the rack in the AO lab). The readout electronics are installed as server blades into a crate that supplies power, 1 PPS, and 10 MHz signals to each set of boards (2 board sets per blade). Total power consumption is dominated by this readout crate that contributes 1.4 kW of the 1.7 kW total power budget. Cooling of the rack is achieved with two fan tray/heat exchanger pairs, one dedicated to the crate and one to the rest of the rack components, integrated with the P3K glycol cooling system through a copper manifold in the back of the rack. Total weight is roughly 100 kg.

\section{Laboratory Verification}
\label{sec:verification}

\subsection{MKID Sensitivity and Energy Resolution}
\label{sec:sensitivity}
\begin{figure}[t]
\begin{center}
\includegraphics[width=0.85\linewidth]{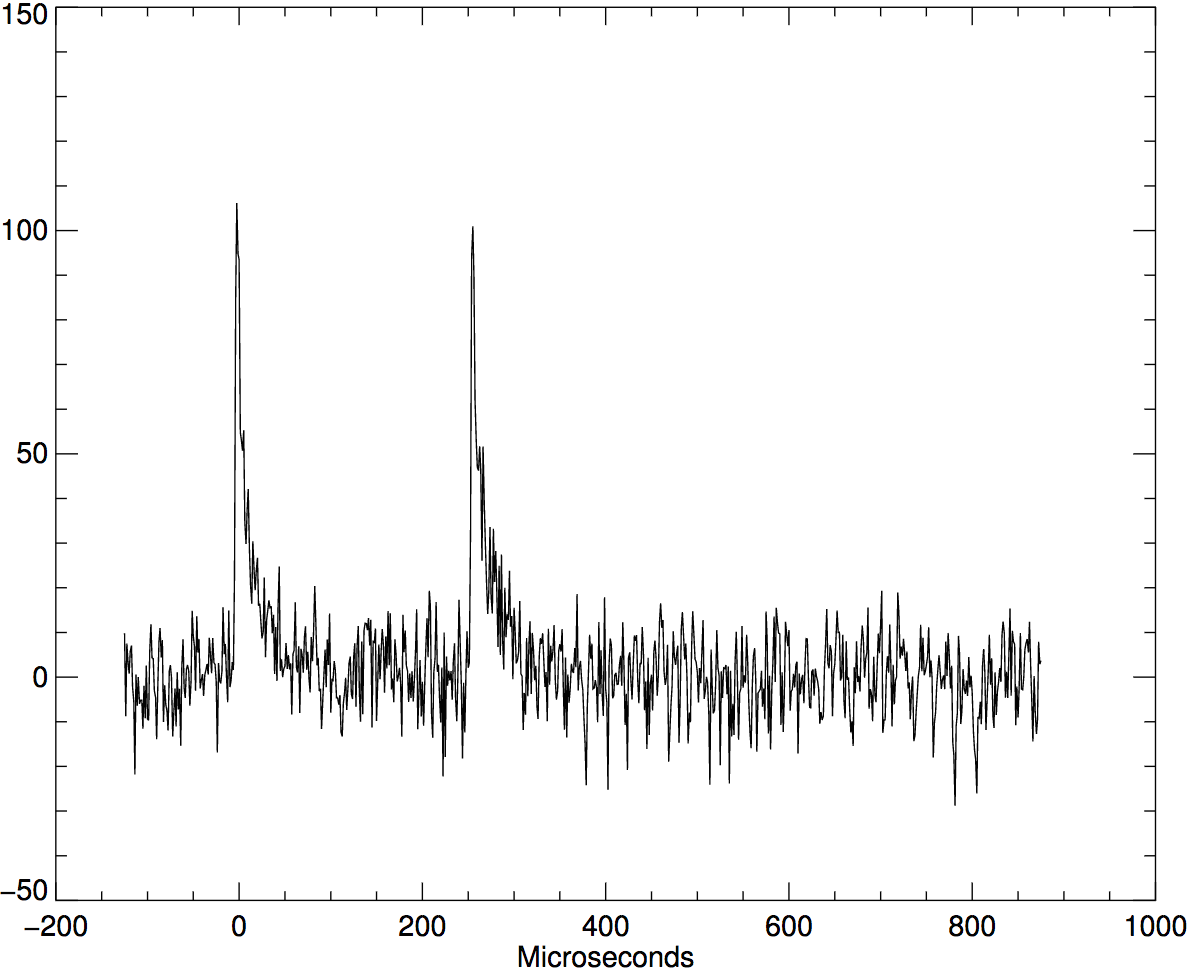}
\end{center}
\caption{Typical D-3 808 nm pulses.} \label{fig:d3pulses}
\end{figure}

To confirm the sensitivity of the detectors we illuminate the array with a laser source in the lab while reading out a handful of resonators with a small-scale version of our readout. Figure~\ref{fig:d3pulses} shows a couple events from an 808~nm laser. The phase offsets are near the desired value of $\sim 100^{\circ}$ for the blue end of our bandpass.

\begin{figure}[t]
\begin{center}
\includegraphics[width=1.\linewidth]{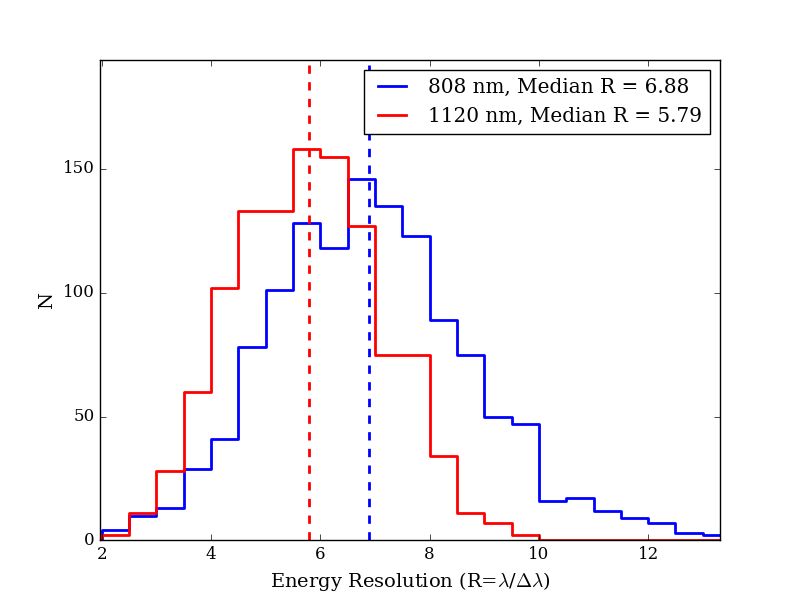}
\end{center}
\caption{Histogram of individual D-3 pixel energy resolutions.} \label{fig:Rhist}
\end{figure}

To calibrate the spectral resolution across the entire array we illuminate with a series of narrow band lasers at different wavelengths. For a given wavelength, $\lambda$, we estimate a pixel's spectral resolution, $R = \lambda/\Delta\lambda$, by the full-width-half-maximum (FWHM) of a Gaussian fit to the pixel's measured spectrum. Using our full digital readout we determine $R$ as a function of $\lambda$ for every pixel. Figure~\ref{fig:Rhist} shows the distribution of measured $R$ from our state-of-the art D-3 array for pixels that received full phase-to-wavelength solutions. As shown in \citet{Szypryt2017}, the current PtSi D-3 arrays exhibit large fluctuations in $Q_{i}$ with resonator frequency, resulting in a significant number of pixels having too low energy resolution to be calibrated in the manner described in Section \ref{sec:wvlcal}.

From this data we see that the median spectral resolution across our 0.8~to~1.4~$\mu$m band goes from $\sim$7 to 5, roughly equivalent in width to the standard near-IR photometric filters at these wavelengths (i.e. Y, J). The current energy resolution is consistent with the measured phase noise in our resonators, dominated by HEMT noise \citep{Mazin2012} and two-level system (TLS) noise.\footnote{TLSs are tunneling states in amorphous solids, and appear at the substrate-superconductor interface in MKIDs even when depositing the superconducting film directly on a clean substrate. An in-depth investigation of this noise source can be found in \citet{Gao2008,GaoThesis}.} Further development is required to improve energy resolution through higher quality resonator fabrication (see \citet{Szypryt2017}) and lower noise-temperature amplifiers.


\subsection{DARKNESS Throughput}
\label{sec:qe}
To measure the absolute throughput of DARKNESS including the 300~K, 77~K, and 4~K windows, MLA, and MKID QE, we've constructed a quantum efficiency (QE) testbed. This testbed uses a monochromator and integrating sphere to generate a uniformly illuminated object plane of given $\lambda$. We then re-image this object plane to a $\sim f$/300 beam and use a rotating fold mirror to send the light out to DARKNESS, or to a calibrated photo-diode inside the testbed enclosure. By dividing the flux measured at DARKNESS by the flux measured on the photo-diode we obtain an absolute measurement of the instrument throughput at each MKID pixel, and with the monochromator we can perform this measurement at discrete wavelengths across our bandpass.

\begin{figure}[t]
\begin{center}
\includegraphics[width=1.\linewidth]{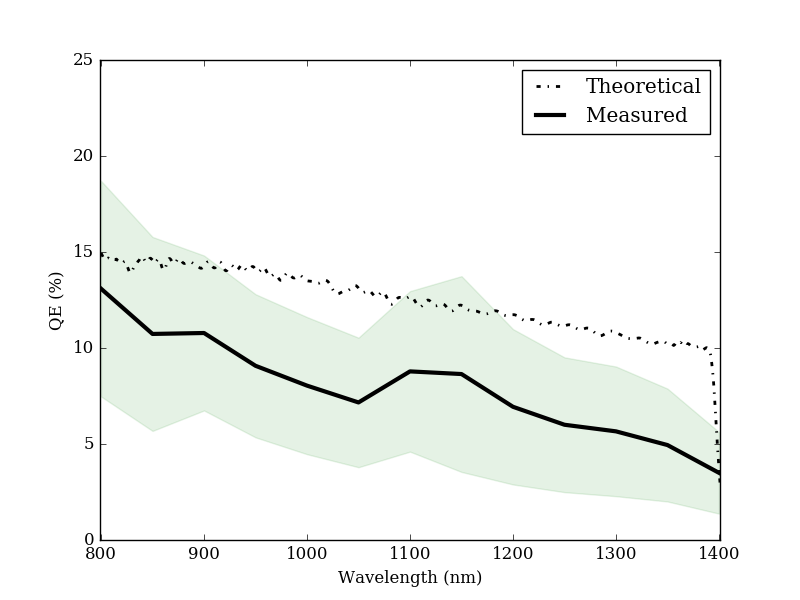}
\end{center}
\caption{DARKNESS measured throughput vs. wavelength compared against theoretical prediction. The solid curve and shaded region shows the median and 1$\sigma$ spread in throughput measured by all pixels. For the theoretical curves we've assumed flat 93\% transmission through the 300~K window, applied the manufacturer transmission curves for the filters at 77~K and 4~K (roughly 97\% across our band), assumed MLA fill factor of 93\% and transmission of 98\% (from manufacturer), inductor fill factor of 90\% due to gaps between the meandered line, and PtSi measured absorptivity from \citet{Szypryt2016}. These parameters are all measured or otherwise well constrained, leaving the MLA spot size as the final factor. From Zemax simulation we expect ensquared energy at the inductor to be 80\%, but this is very sensitive to MLA focus and alignment.} \label{fig:qe}
\end{figure}

Figure~\ref{fig:qe} shows the median throughput measured across the array in 50~nm steps from 0.8 to 1.4~$\mu$m (solid black curve). The shaded region is the 1-$\sigma$ standard deviation in throughput measured by the individual pixels. This measured curve is compared against the throughput expected from the known transmission of our windows + filters, PtSi measured absorptivity \citep{Szypryt2016}, and D-3 inductor fill factor. The biggest unknown in this theoretical curve is the spot size at the MLA focus. From Zemax diffraction simulations we assume 80\% ensquared energy at the inductor, but this still does not account for all the lost flux. We attribute the remaining lost flux to MLA misalignment (roughly 20~$\mu$m of lateral MLA misalignment relative to the MKID inductor centers could account for a factor of 2 loss in QE) and are investigating strategies to improve our MLA mounting precision.

The above throughput calculation is for DARKNESS only. To assess the total capabilities of the instrument we estimate the throughputs of P3K and the SDC to be roughly 50$\%$\footnote{\citet{Dekany2013} estimate photodetection efficiency at the P3K WFS as 24$\%$, and this value was measured as 21$\%$ in 2015 (R. Burruss, private communication). Assuming a conservative 92$\%$ average efficiency per optic, this corresponds to $\sim$50$\%$ final transmission of the science beam at the dichroic.} and 80$\%$ respectively (M. Bottom 2018, private communication).

\section{Commissioning}
\label{sec:commissioning}
DARKNESS traveled to Palomar Observatory for commissioning in July 2016, then again for ongoing commissioning and science verification in November 2016, April 2017, and October 2017. First-light was achieved on July 26, 2016, marking the first demonstration of J-band imaging with an MKID array on sky and the first diffraction limited images obtained with a UVOIR MKID on sky, and April 2017 marked the first deployment of a PtSi MKID array on sky.

\subsection{Data Reduction}
\label{sec:pipeline}
DARKNESS's data reduction is largely based on the ARCONS pipeline \citep{vanEyken2015}, however the fully integrated pipeline to produce final calibrated photon lists is not yet available for DARKNESS data. Development effort so far has been focused on speeding up the initial accessing of the binary data. The ARCONS pipeline is written around the HDF5 file format (H5), and ARCONS data is saved directly to H5 files as a list of photon packets organized by pixel for a user defined total exposure time. The DARKNESS pipeline retains the use of the H5 format for its intermediate files, however, DARKNESS data is communicated to the control computer continuously every 500 $\mu$s and recorded to raw binary files every second as a timestream of photon packets. The first step in the reduction pipeline is to collect the desired 1-second binary files for a given target and format them into an indexed H5 file that can be funneled into the subsequent modules. After that, calibrations are carried out following the procedure in \citet{vanEyken2015} and we perform additional processing and analysis (image registration, photometry, etc.) with a combination of custom \emph{Python} scripts (mostly \emph{NumPy} and \emph{SciPy}) and the Vortex Image Processing pipeline \citep[VIP; ][]{Gomez2017}. Ultimately, these integrated packages will be streamlined and serve as the foundation for our statistical speckle suppression pipeline, to be presented in future work.

The only major deviation from the calibration procedures in \citet{vanEyken2015} is how we perform hot pixel masking. As mentioned in Section~\ref{sec:d3} PtSi on Sapphire resonators do not show the same random ``switching'' behavior seen in our TiN on Silicon arrays. Hot pixels are now mostly related to non-ideal readout parameters, manifesting as either constantly high count rates (bad phase threshold) or ``beating'' in intensity (two readout tones too close together). We simply flag these bad pixels using periodic dark exposures.

To facilitate quicker verification of our on-sky imaging performance, we performed most commissioning observations through a J-band filter. For the results presented in the remainder of this section we forego the spectral calibrations and treat this data as conventional imaging. The photon time streams are binned in time to the desired short exposure time, and these frames are then dark subtracted, flat corrected, aligned, and smoothed with a Gaussian kernel having $\lambda$/D FWHM. 

\subsection{Plate Scale Verification}
\label{sec:results}
\begin{figure}[t]
\begin{center}
\includegraphics[width=0.7\linewidth]{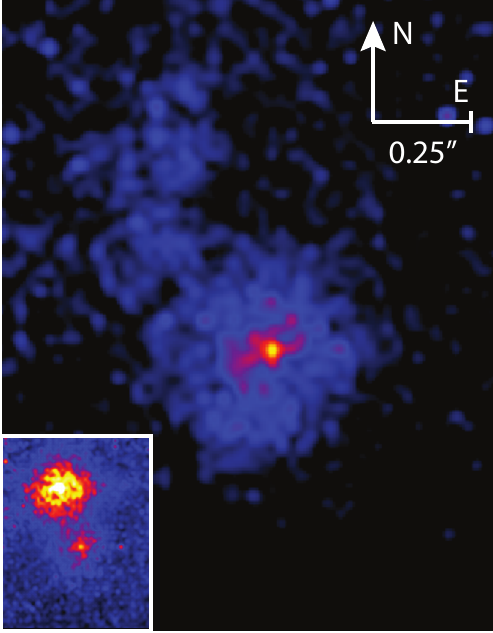}
\end{center}
\caption{Median 1-second J-band image of 10 Uma, a spectroscopic binary with separation of 0.42" at the time of observation, V$_{prim}$=3.96, and $\Delta$V$\approx$2. The large frame shows the system with the coronagraph FPM installed, and inset shows FPM removed to reveal the primary.} \label{fig:commissioning}
\end{figure}

In November 2016 we observed 10 Ursae Majoris (10 Uma), a spectroscopic binary with separation of 0.42" at the time of observation, V$_{prim}$=3.96, and $\Delta$V$\approx$2. Figure~\ref{fig:commissioning} shows a J-band image of the system with the coronagraph FPM installed, and inset shows FPM removed to reveal the primary. We fit a centroid to both objects, and using the known separation we calculate an on-sky plate scale of 19.8~mas per pixel. The diffraction limit $\lambda/D$ for a 5.1~m telescope at 1.25~$\mu$m is roughly 50~mas, so we achieve Nyquist sampling at these wavelengths, but are slightly sub-Nyquist at 0.8~$\mu$m.

\subsection{Raw Contrast}
\label{sec:results}

\begin{figure*}[]
\begin{center}
\includegraphics[width=0.8\textwidth]{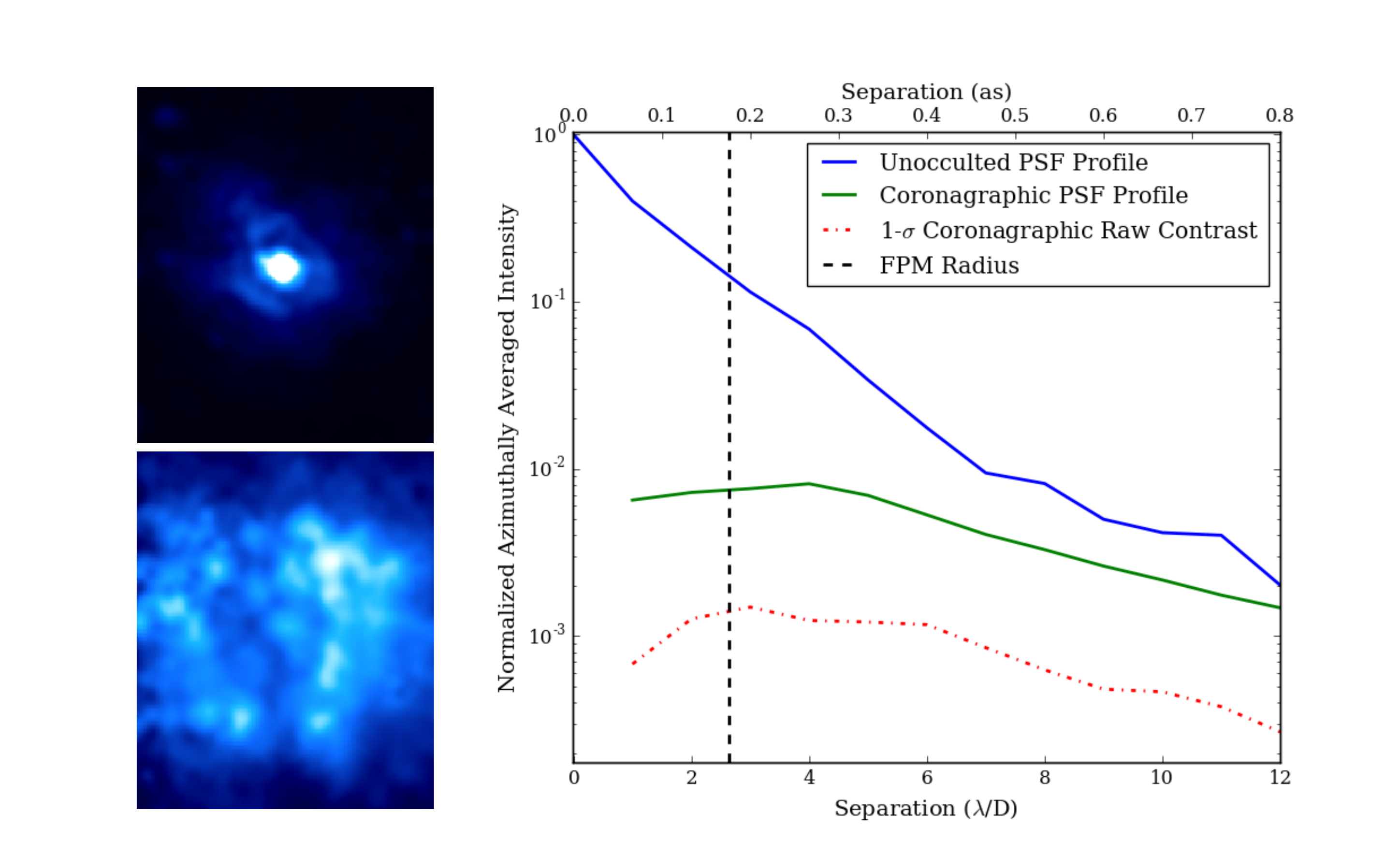}
\end{center}
\caption{Raw J-band contrast (no post-processing speckle subtraction) as a function of angular separation with and without coronagraph FPM installed, measured on-sky with $\pi$ Herculis (J=0.79). (Top Left) Median of 124 one-second frames of the unocculted, unsaturated PSF. (Bottom Left) Median of 1350 one-second frames of the coronagraphic PSF. Both images are \emph{linear} scaled, but with different intensity minima and maxima. (Right) Azimuthally averaged PSF profiles and 1-$\sigma$ contrast vs. separation curve measured in annuli at $\lambda/D$ spacing and normalized by the unocculted, unsaturated PSF core. \emph{Note:} This data was taken with the 20\% undersized Lyot stop, which corresponds to $\lambda/D$ = 63 mas in J-band.} \label{fig:pihercontrast}
\end{figure*}

To estimate the raw contrast achieved with the coronagraph we use observations of $\pi$ Herculis (J=0.79) from April 2017, shown in Figure~\ref{fig:pihercontrast}. We first observed the unocculted PSF using our OD2.5 neutral density (ND) filter to ensure we could perform photometry on an unsaturated PSF core, and also with the coronagraph Lyot stop installed to ensure proper normalization of the throughput. We then moved the coronagraph FPM in and removed the ND filter to observe the surrounding speckle pattern. We proceed using tools from VIP to perform basic aperture photometry on the unocculted PSF and estimate contrast as a function of angular separation. In the standard procedure, standard deviation at a given separation is estimated with a ring of $\lambda/D$ sized apertures. These intensities are then normalized by the intensity of the unocculted, unsaturated PSF core measured with the same aperture. Figure~\ref{fig:pihercontrast} (Left) shows the un-occulted and coronagraphic PSFs. Figure~\ref{fig:pihercontrast} (Right) shows the resulting raw contrast curves (i.e. no post-processing to remove the static speckle pattern), plotting azimuthally averaged PSF profiles and 1-$\sigma$ standard deviation measured in 1$\lambda/D$ annuli. 

We see the coronagraph provides little suppression as there is not significant diffraction to suppress in this low Strehl regime, but we are optimistic that this will improve with P3K's impending upgrade.

\subsection{Preliminary Speckle Statistics Results}
\label{sec:lifetimes}
\begin{figure}[]
\begin{center}
\includegraphics[width=0.75\linewidth]{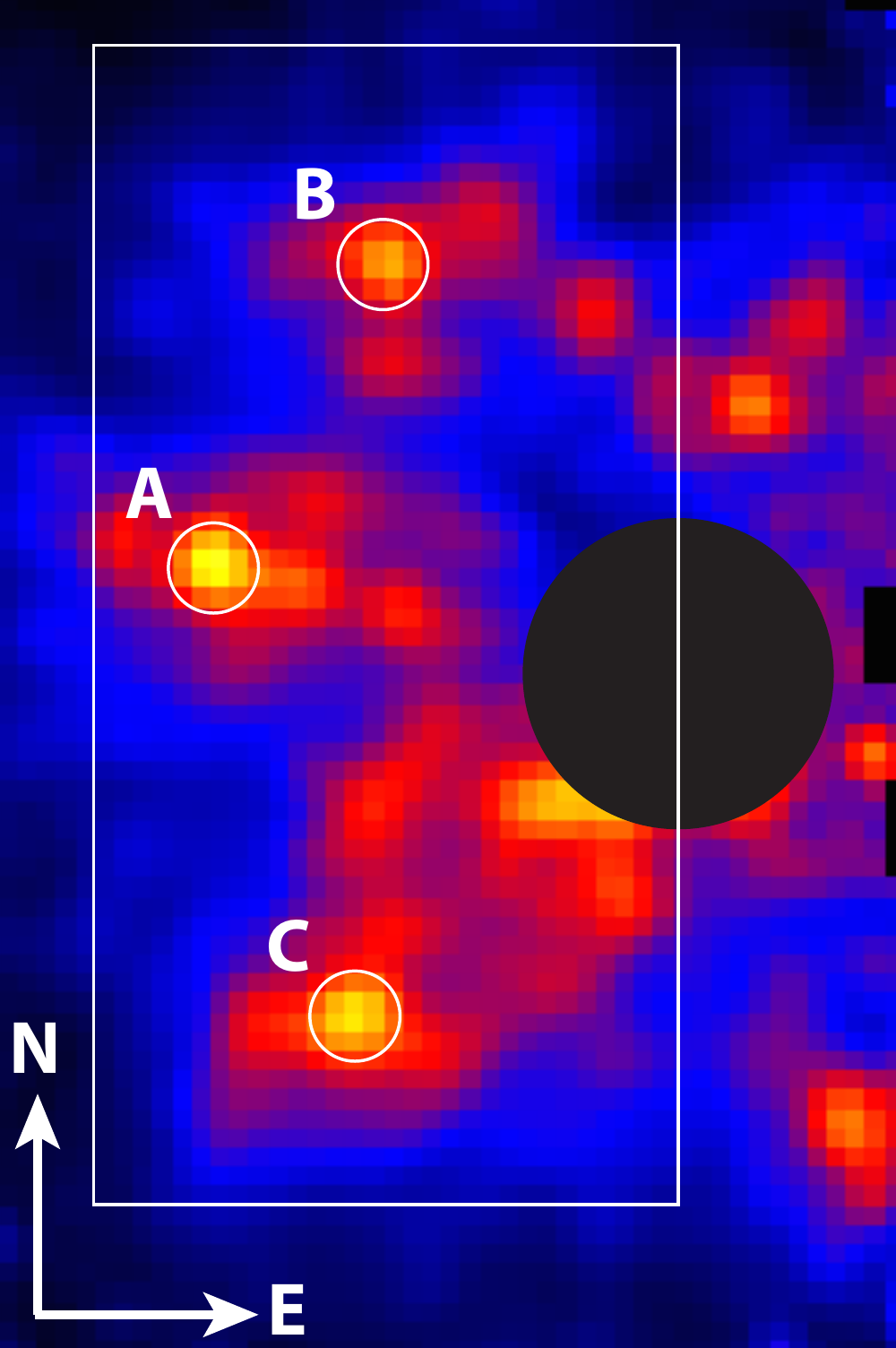}
\end{center}
\caption{Median 1-second J-band image of 32 Peg Ab (at location A) and satellite speckles generated by placing two sine wave shapes on the P3K DM (locations B, C). The white box demarcates the region shown in Figure~\ref{fig:icis_map}.} \label{fig:32peg}
\end{figure}

Here we preview an investigation of speckle vs. companion statistics at very short timescales to demonstrate the promise of MKIDs. On October 4, 2017 we observed the multiple star system 32 Pegasi, of which the closest companion (32 Peg Ab, a stellar companion despite the lower-case "b" in the name) has 0.4" separation and $\Delta$V=4.03 \citep{Mason2001}.\footnote{Unfortunately the data available on Vizier for this object only provides the known visual magnitudes of the two components of the system, and only the spectral type of the primary, precluding an extrapolation of $\Delta$J.} Figure~\ref{fig:32peg} shows the median J-band image from a set of 1-second effective exposures with the coronagraph FPM installed. 32~Peg~Ab is marked at location A, and by adding two sine-waves to the DM shape we have placed satellite speckles (copies of the stellar PSF) of similar intensity to 32~Peg~Ab at locations B and C.

\begin{figure}[]
\begin{center}
\includegraphics[width=1.\linewidth]{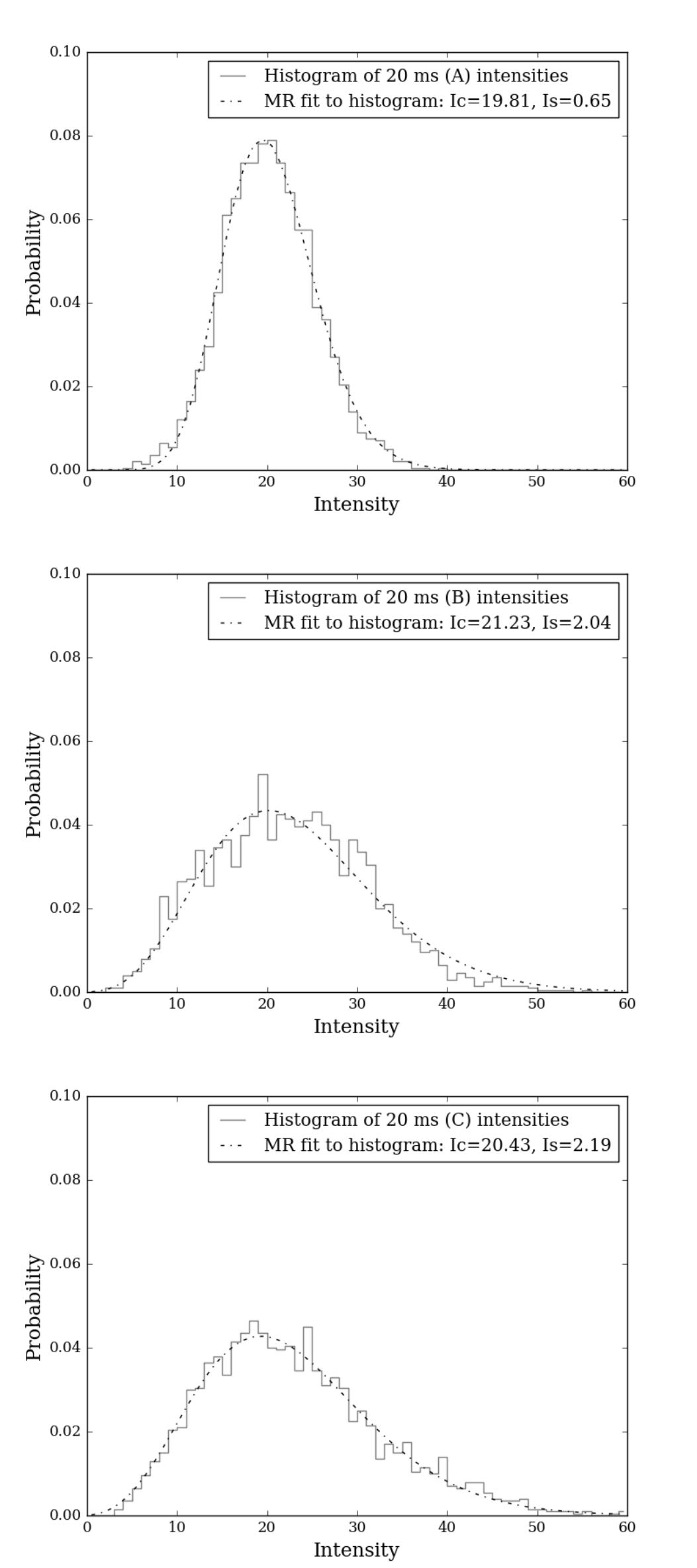}
\end{center}
\caption{Intensity histograms from two-thousand 20~ms exposures for 32 Peg Ab (top panel) and satellite speckles (middle and bottom panels). Intensity units are photon counts in each aperture per 20 ms frame.} \label{fig:32peg_hists}
\end{figure}

With this dataset we show proof-of-concept of the SSD technique. \citet{Gladysz2008} point out that observers should avoid simply using the shortest possible exposure, as low count rates will tend toward Poissonian statistics. On the other hand, too long of an exposure could allow many realizations of the speckle pattern to average together, and by the central limit theorem the MR statistics will become Gaussian. With the power of MKIDs we can dynamically select this exposure time in post-processing. \citet{Macintosh2005} find that atmospheric speckles evolve on timescales related to the aperture clearing time, $\tau_{0} = 0.6~ D/\bar{v}$, where $D$ is the telescope aperture diameter, and $\bar{v}$ is the mean wind speed. With $D$~=~5.1~m for Palomar's Hale telescope, and a prevailing wind speed of 9 m/s at the time of observation, this lifetime is approximately 340~ms. Using a 40~second subset of the full 32~Peg dataset we bin the photon timestream to a conservative 20~ms effective exposures, and plot the resulting histograms of measured intensity at locations A, B, and C, shown in Figure~\ref{fig:32peg_hists}.

\begin{figure}[]
\begin{center}
\includegraphics[width=1.\linewidth]{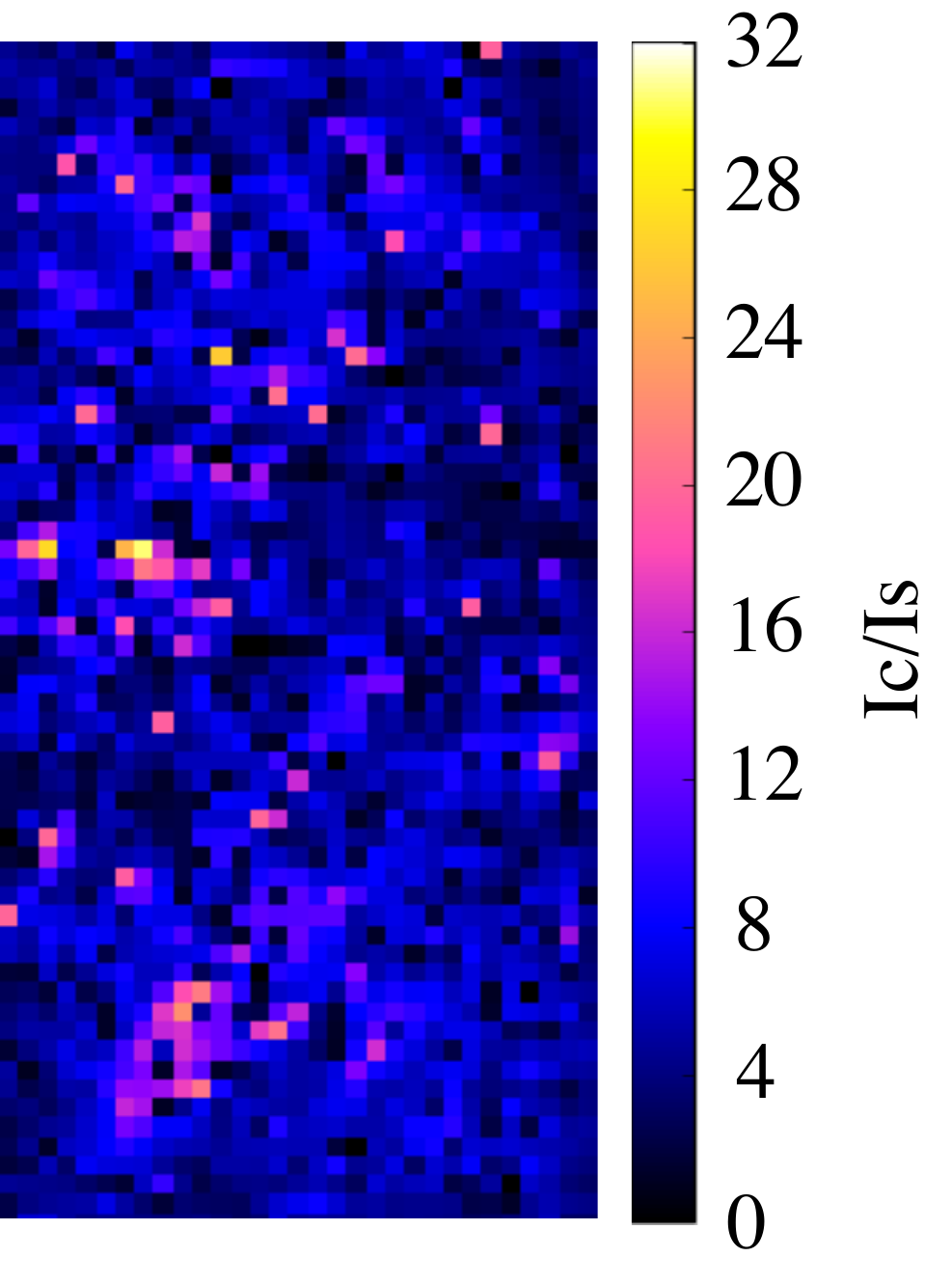}
\end{center}
\caption{Map of $I_{c}/I_{s}$ for each pixel.} \label{fig:icis_map}
\end{figure}

Qualitatively the distribution shape from location A is noticeably different from those of the speckles. Taking inspiration from \citet{Fitzgerald2006} we fit the histograms with MR functions to infer values of $I_{c}$ and $I_{s}$ at each location. We then use the ratio $I_{c}/I_{s}$ as a proxy for each histogram's skewness, and as a means to quantify the differences. Though all three distributions can be fit by MR curves, $I_{c}/I_{s} \approx$~30 at location A and $\approx$~10 at locations B and C. If we make a map of $I_{c}/I_{s}$ at every pixel (shown in Figure~\ref{fig:icis_map}) the effect is clear. With a maximum $I_{c}/I_{s}$ of 30.7 at the center of the companion PSF compared to a mean across the map of 5.8 with SD = 3.8, the companion stands out in this parameter despite being indistinguishable from the speckles in average intensity. A full analysis of the SNR gain from this test is beyond the scope of the current manuscript, but future work will investigate the performance of this and other statistical tests in maximizing companion SNR, with the ultimate goal of generalizing to use the photon arrival time statistics without the need for binning.

\section{Conclusion and On-going Work}
\label{sec:conclusions}
We have presented the design and initial on-sky performance of DARKNESS, a MKID based IFS for high contrast imaging at Palomar. UVOIR MKID technology is still maturing rapidly --- we have active research programs dedicated to improving the PtSi resonator quality factor, which will improve raw pixel yield and energy resolution, as well as exploring direct deposition of anti-reflection coating on the MKID inductors to drastically improve detector QE. The results shown here can be considered as a snapshot of the current state of the instrument, which will be upgraded with a new MKID array when these improvements come online.

The SDC has demonstrated competitive contrasts with speckle nulling \citep{Bottom2016a}, however, the current bottleneck is the long readout time of the science camera, PHARO \citep{Hayward2001}, resulting in slow convergence and limited speckle control due to their decorrelation timescales being comparable to the control loop delays. We are currently working with the P3K team to implement a faster communication scheme between the P3K and DARKNESS control computers, and aim to demonstrate speckle nulling on-sky at $>10$~Hz rates in the coming months (Fruitwalla et al. in prep.).

The WFS camera and real-time controller (RTC) in P3K will be upgraded in early 2018. With these upgrades in place we anticipate improved J-band Strehl ratio and have procured a J-band VVC for use with DARKNESS. The RTC upgrade will also open the door for kHz feedback from outside the main P3K control loop, enabling nulling of atmospheric speckles. With these upgrades in place we expect to overcome the current speckle noise barrier and achieve photon-noise limited contrasts.

DARKNESS represents the beginning of a multi-instrument effort to utilize MKIDs for exoplanet imaging. The development work invested in DARKNESS has simultaneously supported the MKID Exoplanet Camera (MEC), a 20~kilopixel MKID IFS for SCExAO shipping to Subaru in late 2017, and PICTURE-C \citep{Cook2015}, a balloon-borne high-contrast platform that will fly a DARKNESS clone in 2019. Additionally, thanks to its portable design, DARKNESS is slated to join MagAO-X \citep{Males2016} in 2019 as the first high-contrast instrument with MKID IFS backend in the Southern Hemisphere.

\acknowledgments
The authors would like to thank the anonymous referee for their thorough comments. We also extend a special thanks to the amazing staff of Palomar Observatory. Their patience, knowledge, and support have been vital to the successful design and operation of DARKNESS.

DARKNESS was funded by an NSF ATI grant AST-1308556. SRM, PS, and NZ were supported throughout this work by NASA Office of the Chief Technologist's Space Technology Research Fellowships (NSTRF). The research was carried out in part at the Jet Propulsion Laboratory, California Institute of Technology, under a contract with the National Aeronautics and Space Administration. This work is based on observations obtained at the Hale Telescope, Palomar Observatory as part of a continuing collaboration between the California Institute of Technology, NASA/JPL, Yale University, and the National Astronomical Observatories of China

\facilities{Hale (PALM-3000, Stellar Double Coronagraph)} 

\newpage
\bibliographystyle{yahapj}
\bibliography{ms}

\begin{thebibliography}{}
\providecommand\natexlab[1]{#1}
\providecommand\JournalTitle[1]{#1}

\bibitem[{{Aime} \& {Soummer}(2004)}]{Aime2004}
{Aime}, C., \& {Soummer}, R. 2004,
  \href{http://dx.doi.org/10.1086/424381}{\JournalTitle{ApJL}, 612, L85}

\bibitem[{{Boccaletti} {et~al.}(1998){Boccaletti}, {Labeyrie}, \&
  {Ragazzoni}}]{Boccaletti1998}
{Boccaletti}, A., {Labeyrie}, A., \& {Ragazzoni}, R. 1998,
  \JournalTitle{A\&AS}, 338, 106

\bibitem[{{Boccaletti} {et~al.}(2001){Boccaletti}, {Moutou}, {Mouillet},
  {Lagrange}, \& {Augereau}}]{Boccaletti2001}
{Boccaletti}, A., {Moutou}, C., {Mouillet}, D., {Lagrange}, A.-M., \&
  {Augereau}, J.-C. 2001,
  \href{http://dx.doi.org/10.1051/0004-6361:20000342}{\JournalTitle{A\&A}, 367,
  371}

\bibitem[{{Bord{\'e}} \& {Traub}(2006)}]{Borde&Traub}
{Bord{\'e}}, P.~J., \& {Traub}, W.~A. 2006,
  \href{http://dx.doi.org/10.1086/498669}{\JournalTitle{ApJ}, 638, 488}

\bibitem[{{Bottom} {et~al.}(2016{\natexlab{a}}){Bottom}, {Femenia}, {Huby},
  {Mawet}, {Dekany}, {Milburn}, \& {Serabyn}}]{Bottom2016b}
{Bottom}, M., {Femenia}, B., {Huby}, E., {et~al.} 2016{\natexlab{a}},
  \href{http://dx.doi.org/10.1117/12.2233025}{in Proc. SPIE, Vol. 9909,
  Adaptive Optics Systems V}, 990955

\bibitem[{{Bottom} {et~al.}(2016{\natexlab{b}}){Bottom}, {Shelton}, {Wallace},
  {Bartos}, {Kuhn}, {Mawet}, {Mennesson}, {Burruss}, \&
  {Serabyn}}]{Bottom2016a}
{Bottom}, M., {Shelton}, J.~C., {Wallace}, J.~K., {et~al.} 2016{\natexlab{b}},
  \href{http://dx.doi.org/10.1088/1538-3873/128/965/075003}{\JournalTitle{PASP},
  128, 075003}

\bibitem[{{Bowler}(2016)}]{Bowler2016}
{Bowler}, B.~P. 2016,
  \href{http://dx.doi.org/10.1088/1538-3873/128/968/102001}{\JournalTitle{PASP},
  128, 102001}

\bibitem[{{Burruss} {et~al.}(2010){Burruss}, {Serabyn}, {Mawet}, {Roberts},
  {Hickey}, {Rykoski}, {Bikkannavar}, \& {Crepp}}]{Burruss2010}
{Burruss}, R.~S., {Serabyn}, E., {Mawet}, D.~P., {et~al.} 2010,
  \href{http://dx.doi.org/10.1117/12.857544}{in \procspie, Vol. 7736, Adaptive
  Optics Systems II}, 77365X

\bibitem[{{Cady} {et~al.}(2013){Cady}, {Baranec}, {Beichman}, {Brenner},
  {Burruss}, {Crepp}, {Dekany}, {Hale}, {Hillenbrand}, {Hinkley}, {Ligon},
  {Lockhart}, {Oppenheimer}, {Parry}, {Pueyo}, {Rice}, {Roberts}, {Roberts},
  {Shao}, {Sivaramakrishnan}, {Soummer}, {Tang}, {Truong}, {Vasisht},
  {Vescelus}, {Wallace}, {Zhai}, \& {Zimmerman}}]{Cady2013}
{Cady}, E., {Baranec}, C., {Beichman}, C., {et~al.} 2013, in ,
  \href{http://arxiv.org/abs/1309.6357}{{\sffamily arXiv:1309.6357
  [astro-ph.IM]}}

\bibitem[{{Claudi} {et~al.}(2008){Claudi}, {Turatto}, {Gratton}, {Antichi},
  {Bonavita}, {Bruno}, {Cascone}, {De Caprio}, {Desidera}, {Giro}, {Mesa},
  {Scuderi}, {Dohlen}, {Beuzit}, \& {Puget}}]{Claudi2008}
{Claudi}, R.~U., {Turatto}, M., {Gratton}, R.~G., {et~al.} 2008,
  \href{http://dx.doi.org/10.1117/12.788366}{in }, 70143E

\bibitem[{{Collura} {et~al.}(2017){Collura}, {Strader}, {Meeker}, {Szypryt},
  {Walter}, {Bockstiegel}, {Mazin}, \& {Prince}}]{Collura2017}
{Collura}, G., {Strader}, P., {Meeker}, S.~R., {et~al.} 2017,
  \href{http://dx.doi.org/10.3847/1538-4357/aa93d8}{\JournalTitle{\apj}, 850,
  65}

\bibitem[{{Cook} {et~al.}(2015){Cook}, {Cahoy}, {Chakrabarti}, {Douglas},
  {Finn}, {Kuchner}, {Lewis}, {Marinan}, {Martel}, {Mawet}, {Mazin}, {Meeker},
  {Mendillo}, {Serabyn}, {Stuchlik}, \& {Swain}}]{Cook2015}
{Cook}, T., {Cahoy}, K., {Chakrabarti}, S., {et~al.} 2015,
  \href{http://dx.doi.org/10.1117/1.JATIS.1.4.044001}{\JournalTitle{JATIS}, 1,
  044001}

\bibitem[{{Crepp} {et~al.}(2011){Crepp}, {Pueyo}, {Brenner}, {Oppenheimer},
  {Zimmerman}, {Hinkley}, {Parry}, {King}, {Vasisht}, {Beichman},
  {Hillenbrand}, {Dekany}, {Shao}, {Burruss}, {Roberts}, {Bouchez}, {Roberts},
  \& {Soummer}}]{Crepp2011}
{Crepp}, J.~R., {Pueyo}, L., {Brenner}, D., {et~al.} 2011,
  \href{http://dx.doi.org/10.1088/0004-637X/729/2/132}{\JournalTitle{ApJ}, 729,
  132}

\bibitem[{{Day} {et~al.}(2003){Day}, {LeDuc}, {Mazin}, {Vayonakis}, \&
  {Zmuidzinas}}]{Day2003}
{Day}, P.~K., {LeDuc}, H.~G., {Mazin}, B.~A., {Vayonakis}, A., \& {Zmuidzinas},
  J. 2003, \href{http://dx.doi.org/10.1038/nature02037}{\JournalTitle{Nature},
  425, 817}

\bibitem[{{Dekany} {et~al.}(2013){Dekany}, {Roberts}, {Burruss}, {Bouchez},
  {Truong}, {Baranec}, {Guiwits}, {Hale}, {Angione}, {Trinh}, {Zolkower},
  {Shelton}, {Palmer}, {Henning}, {Croner}, {Troy}, {McKenna}, {Tesch},
  {Hildebrandt}, \& {Milburn}}]{Dekany2013}
{Dekany}, R., {Roberts}, J., {Burruss}, R., {et~al.} 2013,
  \href{http://dx.doi.org/10.1088/0004-637X/776/2/130}{\JournalTitle{ApJ}, 776,
  130}

\bibitem[{{Dohlen} {et~al.}(2006){Dohlen}, {Beuzit}, {Feldt}, {Mouillet},
  {Puget}, {Antichi}, {Baruffolo}, {Baudoz}, {Berton}, {Boccaletti},
  {Carbillet}, {Charton}, {Claudi}, {Downing}, {Fabron}, {Feautrier},
  {Fedrigo}, {Fusco}, {Gach}, {Gratton}, {Hubin}, {Kasper}, {Langlois},
  {Longmore}, {Moutou}, {Petit}, {Pragt}, {Rabou}, {Rousset}, {Saisse},
  {Schmid}, {Stadler}, {Stamm}, {Turatto}, {Waters}, \& {Wildi}}]{Dohlen2006}
{Dohlen}, K., {Beuzit}, J.-L., {Feldt}, M., {et~al.} 2006,
  \href{http://dx.doi.org/10.1117/12.671537}{in }

\bibitem[{{Fitzgerald} \& {Graham}(2006)}]{Fitzgerald2006}
{Fitzgerald}, M.~P., \& {Graham}, J.~R. 2006,
  \href{http://dx.doi.org/10.1086/498339}{\JournalTitle{ApJ}, 637, 541}

\bibitem[{{Gao}(2008)}]{GaoThesis}
{Gao}, J. 2008, PhD thesis, California Institute of Technology

\bibitem[{{Gao} {et~al.}(2008){Gao}, {Daal}, {Vayonakis}, {Kumar},
  {Zmuidzinas}, {Sadoulet}, {Mazin}, {Day}, \& {Leduc}}]{Gao2008}
{Gao}, J., {Daal}, M., {Vayonakis}, A., {et~al.} 2008,
  \href{http://dx.doi.org/10.1063/1.2906373}{\JournalTitle{Applied Physics
  Letters}, 92, 152505}

\bibitem[{{Gladysz} \& {Christou}(2008)}]{Gladysz2008}
{Gladysz}, S., \& {Christou}, J.~C. 2008,
  \href{http://dx.doi.org/10.1086/589679}{\JournalTitle{ApJ}, 684, 1486}

\bibitem[{{Gladysz} \& {Christou}(2009)}]{Gladysz2009}
---. 2009,
  \href{http://dx.doi.org/10.1088/0004-637X/698/1/28}{\JournalTitle{ApJ}, 698,
  28}

\bibitem[{{Gomez Gonzalez} {et~al.}(2017){Gomez Gonzalez}, {Wertz}, {Absil},
  {Christiaens}, {Defr{\`e}re}, {Mawet}, {Milli}, {Absil}, {Van Droogenbroeck},
  {Cantalloube}, {Hinz}, {Skemer}, {Karlsson}, \& {Surdej}}]{Gomez2017}
{Gomez Gonzalez}, C.~A., {Wertz}, O., {Absil}, O., {et~al.} 2017,
  \href{http://dx.doi.org/10.3847/1538-3881/aa73d7}{\JournalTitle{AJ}, 154, 7}

\bibitem[{{Goodman}(2005)}]{Goodman2005}
{Goodman}, J.~W. 2005, {Introduction to Fourier optics}

\bibitem[{{Groff} {et~al.}(2014){Groff}, {Kasdin}, {Limbach}, {Galvin}, {Carr},
  {Knapp}, {Brandt}, {Loomis}, {Jarosik}, {Mede}, {McElwain}, {Janson},
  {Guyon}, {Jovanovic}, {Takato}, {Martinache}, \& {Hayashi}}]{Groff2014}
{Groff}, T.~D., {Kasdin}, N.~J., {Limbach}, M.~A., {et~al.} 2014,
  \href{http://dx.doi.org/10.1117/12.2055769}{in Proc. SPIE, Vol. 9147,
  Ground-based and Airborne Instrumentation for Astronomy V}, 91471W

\bibitem[{{Guyon}(2005)}]{Guyon2005}
{Guyon}, O. 2005, \href{http://dx.doi.org/10.1086/431209}{\JournalTitle{ApJ},
  629, 592}

\bibitem[{{Hayward} {et~al.}(2001){Hayward}, {Brandl}, {Pirger}, {Blacken},
  {Gull}, {Schoenwald}, \& {Houck}}]{Hayward2001}
{Hayward}, T.~L., {Brandl}, B., {Pirger}, B., {et~al.} 2001,
  \href{http://dx.doi.org/10.1086/317969}{\JournalTitle{PASP}, 113, 105}

\bibitem[{{Hinkley} {et~al.}(2008){Hinkley}, {Oppenheimer}, {Brenner}, {Parry},
  {Sivaramakrishnan}, {Soummer}, \& {King}}]{Hinkley2008}
{Hinkley}, S., {Oppenheimer}, B.~R., {Brenner}, D., {et~al.} 2008,
  \href{http://dx.doi.org/10.1117/12.789557}{in }

\bibitem[{{Hinkley} {et~al.}(2007){Hinkley}, {Oppenheimer}, {Soummer},
  {Sivaramakrishnan}, {Roberts}, {Kuhn}, {Makidon}, {Perrin}, {Lloyd},
  {Kratter}, \& {Brenner}}]{Hinkley2007}
{Hinkley}, S., {Oppenheimer}, B.~R., {Soummer}, R., {et~al.} 2007,
  \href{http://dx.doi.org/10.1086/509063}{\JournalTitle{\apj}, 654, 633}

\bibitem[{{Hinkley} {et~al.}(2011){Hinkley}, {Oppenheimer}, {Zimmerman},
  {Brenner}, {Parry}, {Crepp}, {Vasisht}, {Ligon}, {King}, {Soummer},
  {Sivaramakrishnan}, {Beichman}, {Shao}, {Roberts}, {Bouchez}, {Dekany},
  {Pueyo}, {Roberts}, {Lockhart}, {Zhai}, {Shelton}, \&
  {Burruss}}]{Hinkley2011}
{Hinkley}, S., {Oppenheimer}, B.~R., {Zimmerman}, N., {et~al.} 2011,
  \href{http://dx.doi.org/10.1086/658163}{\JournalTitle{PASP}, 123, 74}

\bibitem[{{Jovanovic} {et~al.}(2015){Jovanovic}, {Martinache}, {Guyon},
  {Clergeon}, {Singh}, {Kudo}, {Garrel}, {Newman}, {Doughty}, {Lozi}, {Males},
  {Minowa}, {Hayano}, {Takato}, {Morino}, {Kuhn}, {Serabyn}, {Norris},
  {Tuthill}, {Schworer}, {Stewart}, {Close}, {Huby}, {Perrin}, {Lacour},
  {Gauchet}, {Vievard}, {Murakami}, {Oshiyama}, {Baba}, {Matsuo}, {Nishikawa},
  {Tamura}, {Lai}, {Marchis}, {Duchene}, {Kotani}, \&
  {Woillez}}]{Jovanovic2015}
{Jovanovic}, N., {Martinache}, F., {Guyon}, O., {et~al.} 2015,
  \href{http://dx.doi.org/10.1086/682989}{\JournalTitle{PASP}, 127, 890}

\bibitem[{{Labeyrie}(1995)}]{Labeyrie1995}
{Labeyrie}, A. 1995, \JournalTitle{A\&A}, 298, 544

\bibitem[{{Larkin} {et~al.}(2014){Larkin}, {Chilcote}, {Aliado}, {Bauman},
  {Brims}, {Canfield}, {Cardwell}, {Dillon}, {Doyon}, {Dunn}, {Fitzgerald},
  {Graham}, {Goodsell}, {Hartung}, {Hibon}, {Ingraham}, {Johnson}, {Kress},
  {Konopacky}, {Macintosh}, {Magnone}, {Maire}, {McLean}, {Palmer}, {Perrin},
  {Quiroz}, {Rantakyr{\"o}}, {Sadakuni}, {Saddlemyer}, {Serio}, {Thibault},
  {Thomas}, {Vallee}, \& {Weiss}}]{Larkin2014}
{Larkin}, J.~E., {Chilcote}, J.~K., {Aliado}, T., {et~al.} 2014,
  \href{http://dx.doi.org/10.1117/12.2056504}{in }, 91471K

\bibitem[{{Leduc} {et~al.}(2010){Leduc}, {Bumble}, {Day}, {Eom}, {Gao},
  {Golwala}, {Mazin}, {McHugh}, {Merrill}, {Moore}, {Noroozian}, {Turner}, \&
  {Zmuidzinas}}]{Leduc2010}
{Leduc}, H.~G., {Bumble}, B., {Day}, P.~K., {et~al.} 2010,
  \href{http://dx.doi.org/10.1063/1.3480420}{\JournalTitle{APL}, 97, 102509}

\bibitem[{{Macintosh} {et~al.}(2005){Macintosh}, {Poyneer}, {Sivaramakrishnan},
  \& {Marois}}]{Macintosh2005}
{Macintosh}, B., {Poyneer}, L., {Sivaramakrishnan}, A., \& {Marois}, C. 2005,
  \href{http://dx.doi.org/10.1117/12.627854}{in Proc. SPIE, Vol. 5903,
  Astronomical Adaptive Optics Systems and Applications II, ed. R.~K. {Tyson}
  \& M.~{Lloyd-Hart}}, 170

\bibitem[{{Macintosh} {et~al.}(2006){Macintosh}, {Graham}, {Palmer}, {Doyon},
  {Gavel}, {Larkin}, {Oppenheimer}, {Saddlemyer}, {Wallace}, {Bauman}, {Evans},
  {Erikson}, {Morzinski}, {Phillion}, {Poyneer}, {Sivaramakrishnan}, {Soummer},
  {Thibault}, \& {Veran}}]{Macintosh2006}
{Macintosh}, B., {Graham}, J., {Palmer}, D., {et~al.} 2006,
  \href{http://dx.doi.org/10.1117/12.672430}{in }

\bibitem[{{Macintosh} {et~al.}(2014){Macintosh}, {Graham}, {Ingraham},
  {Konopacky}, {Marois}, {Perrin}, {Poyneer}, {Bauman}, {Barman}, {Burrows},
  {Cardwell}, {Chilcote}, {De Rosa}, {Dillon}, {Doyon}, {Dunn}, {Erikson},
  {Fitzgerald}, {Gavel}, {Goodsell}, {Hartung}, {Hibon}, {Kalas}, {Larkin},
  {Maire}, {Marchis}, {Marley}, {McBride}, {Millar-Blanchaer}, {Morzinski},
  {Norton}, {Oppenheimer}, {Palmer}, {Patience}, {Pueyo}, {Rantakyro},
  {Sadakuni}, {Saddlemyer}, {Savransky}, {Serio}, {Soummer},
  {Sivaramakrishnan}, {Song}, {Thomas}, {Wallace}, {Wiktorowicz}, \&
  {Wolff}}]{Macintosh2014}
{Macintosh}, B., {Graham}, J.~R., {Ingraham}, P., {et~al.} 2014,
  \href{http://dx.doi.org/10.1073/pnas.1304215111}{\JournalTitle{PNAS}, 111,
  12661}

\bibitem[{{Males} {et~al.}(2016){Males}, {Close}, {Guyon}, {Morzinski}, {Hinz},
  {Esposito}, {Pinna}, {Xompero}, {Briguglio}, {Riccardi}, {Puglisi}, {Mazin},
  {Ireland}, {Weinberger}, {Conrad}, {Kenworthy}, {Snik}, {Otten}, {Jovanovic},
  \& {Lozi}}]{Males2016}
{Males}, J.~R., {Close}, L.~M., {Guyon}, O., {et~al.} 2016,
  \href{http://dx.doi.org/10.1117/12.2234105}{in \procspie, Vol. 9909, Adaptive
  Optics Systems V}, 990952

\bibitem[{{Marois} {et~al.}(2006){Marois}, {Lafreni{\`e}re}, {Doyon},
  {Macintosh}, \& {Nadeau}}]{Marois2006}
{Marois}, C., {Lafreni{\`e}re}, D., {Doyon}, R., {Macintosh}, B., \& {Nadeau},
  D. 2006, \href{http://dx.doi.org/10.1086/500401}{\JournalTitle{\apj}, 641,
  556}

\bibitem[{{Marois} {et~al.}(2008){Marois}, {Lafreni{\`e}re}, {Macintosh}, \&
  {Doyon}}]{MaroisCL2008}
{Marois}, C., {Lafreni{\`e}re}, D., {Macintosh}, B., \& {Doyon}, R. 2008,
  \href{http://dx.doi.org/10.1086/523839}{\JournalTitle{ApJ}, 673, 647}

\bibitem[{{Marsden} {et~al.}(2012){Marsden}, {Mazin}, {Bumble}, {Meeker},
  {O'Brien}, {McHugh}, {Strader}, \& {Langman}}]{Marsden2012}
{Marsden}, D., {Mazin}, B.~A., {Bumble}, B., {et~al.} 2012,
  \href{http://dx.doi.org/10.1117/12.924904}{in }

\bibitem[{{Martinache} {et~al.}(2014){Martinache}, {Guyon}, {Jovanovic},
  {Clergeon}, {Singh}, {Kudo}, {Currie}, {Thalmann}, {McElwain}, \&
  {Tamura}}]{Martinache2014}
{Martinache}, F., {Guyon}, O., {Jovanovic}, N., {et~al.} 2014,
  \href{http://dx.doi.org/10.1086/677141}{\JournalTitle{PASP}, 126, 565}

\bibitem[{{Mason} {et~al.}(2001){Mason}, {Wycoff}, {Hartkopf}, {Douglass}, \&
  {Worley}}]{Mason2001}
{Mason}, B.~D., {Wycoff}, G.~L., {Hartkopf}, W.~I., {Douglass}, G.~G., \&
  {Worley}, C.~E. 2001,
  \href{http://dx.doi.org/10.1086/323920}{\JournalTitle{\aj}, 122, 3466}

\bibitem[{Mawet {et~al.}(2005)Mawet, Riaud, Absil, \& Surdej}]{Mawet2005}
Mawet, D., Riaud, P., Absil, O., \& Surdej, J. 2005,
  \href{http://dx.doi.org/10.1086/462409}{\JournalTitle{ApJ}, 633, 1191}

\bibitem[{Mawet {et~al.}(2011)Mawet, Serabyn, Wallace, \&
  Pueyo}]{mawet_improved_2011}
Mawet, D., Serabyn, E., Wallace, J.~K., \& Pueyo, L. 2011,
  \href{http://dx.doi.org/10.1364/OL.36.001506}{\JournalTitle{Optics Letters},
  36, 1506}

\bibitem[{{Mawet} {et~al.}(2014){Mawet}, {Milli}, {Wahhaj}, {Pelat}, {Absil},
  {Delacroix}, {Boccaletti}, {Kasper}, {Kenworthy}, {Marois}, {Mennesson}, \&
  {Pueyo}}]{MawetCL2014}
{Mawet}, D., {Milli}, J., {Wahhaj}, Z., {et~al.} 2014,
  \href{http://dx.doi.org/10.1088/0004-637X/792/2/97}{\JournalTitle{ApJ}, 792,
  97}

\bibitem[{{Mazin} {et~al.}(2012){Mazin}, {Bumble}, {Meeker}, {O'Brien},
  {McHugh}, \& {Langman}}]{Mazin2012}
{Mazin}, B.~A., {Bumble}, B., {Meeker}, S.~R., {et~al.} 2012,
  \href{http://dx.doi.org/10.1364/OE.20.001503}{\JournalTitle{Optics Express},
  20, 1503}

\bibitem[{{Mazin} {et~al.}(2013){Mazin}, {Meeker}, {Strader}, {Szypryt},
  {Marsden}, {van Eyken}, {Duggan}, {Walter}, {Ulbricht}, {Johnson}, {Bumble},
  {O''Brien}, \& {Stoughton}}]{Mazin2013}
{Mazin}, B.~A., {Meeker}, S.~R., {Strader}, M.~J., {et~al.} 2013,
  \href{http://dx.doi.org/10.1086/674013}{\JournalTitle{PASP}, 125, 1348}

\bibitem[{{McHugh} {et~al.}(2012){McHugh}, {Mazin}, {Serfass}, {Meeker},
  {O'Brien}, {Duan}, {Raffanti}, \& {Werthimer}}]{McHugh2012}
{McHugh}, S., {Mazin}, B.~A., {Serfass}, B., {et~al.} 2012,
  \href{http://dx.doi.org/10.1063/1.3700812}{\JournalTitle{Review of Scientific
  Instruments}, 83, 044702}

\bibitem[{{Meeker} {et~al.}(2015){Meeker}, {Mazin}, {Jensen-Clem}, {Walter},
  {Szypryt}, {Strader}, \& {Bockstiegel}}]{Meeker2015}
{Meeker}, S.~R., {Mazin}, B.~A., {Jensen-Clem}, R., {et~al.} 2015, in

\bibitem[{{Oppenheimer} {et~al.}(2013){Oppenheimer}, {Baranec}, {Beichman},
  {Brenner}, {Burruss}, {Cady}, {Crepp}, {Dekany}, {Fergus}, {Hale},
  {Hillenbrand}, {Hinkley}, {Hogg}, {King}, {Ligon}, {Lockhart}, {Nilsson},
  {Parry}, {Pueyo}, {Rice}, {Roberts}, {Roberts}, {Shao}, {Sivaramakrishnan},
  {Soummer}, {Truong}, {Vasisht}, {Veicht}, {Vescelus}, {Wallace}, {Zhai}, \&
  {Zimmerman}}]{Oppenheimer2013}
{Oppenheimer}, B.~R., {Baranec}, C., {Beichman}, C., {et~al.} 2013,
  \href{http://dx.doi.org/10.1088/0004-637X/768/1/24}{\JournalTitle{ApJ}, 768,
  24}

\bibitem[{{Perrin} {et~al.}(2003){Perrin}, {Sivaramakrishnan}, {Makidon},
  {Oppenheimer}, \& {Graham}}]{Perrin2003}
{Perrin}, M.~D., {Sivaramakrishnan}, A., {Makidon}, R.~B., {Oppenheimer},
  B.~R., \& {Graham}, J.~R. 2003,
  \href{http://dx.doi.org/10.1086/377689}{\JournalTitle{\apj}, 596, 702}

\bibitem[{{Soummer} {et~al.}(2007){Soummer}, {Ferrari}, {Aime}, \&
  {Jolissaint}}]{Soummer2007}
{Soummer}, R., {Ferrari}, A., {Aime}, C., \& {Jolissaint}, L. 2007,
  \href{http://dx.doi.org/10.1086/520913}{\JournalTitle{\apj}, 669, 642}

\bibitem[{{Sparks} \& {Ford}(2002)}]{Sparks2002}
{Sparks}, W.~B., \& {Ford}, H.~C. 2002,
  \href{http://dx.doi.org/10.1086/342401}{\JournalTitle{ApJ}, 578, 543}

\bibitem[{{Strader}(2016)}]{StraderThesis}
{Strader}, M.~J. 2016, PhD thesis, University of California Santa Barbara

\bibitem[{{Strader} {et~al.}(2013){Strader}, {Johnson}, {Mazin}, {Spiro
  Jaeger}, {Gwinn}, {Meeker}, {Szypryt}, {van Eyken}, {Marsden}, {O'Brien},
  {Walter}, {Ulbricht}, {Stoughton}, \& {Bumble}}]{Strader2013}
{Strader}, M.~J., {Johnson}, M.~D., {Mazin}, B.~A., {et~al.} 2013,
  \href{http://dx.doi.org/10.1088/2041-8205/779/1/L12}{\JournalTitle{ApJ}, 779,
  L12}

\bibitem[{{Strader} {et~al.}(2016){Strader}, {Archibald}, {Meeker}, {Szypryt},
  {Walter}, {van Eyken}, {Ulbricht}, {Stoughton}, {Bumble}, {Kaplan}, \&
  {Mazin}}]{Strader2016}
{Strader}, M.~J., {Archibald}, A.~M., {Meeker}, S.~R., {et~al.} 2016,
  \href{http://dx.doi.org/10.1093/mnras/stw663}{\JournalTitle{MNRAS}, 459, 427}

\bibitem[{{Szypryt} {et~al.}(2016){Szypryt}, {Mazin}, {Ulbricht}, {Bumble},
  {Meeker}, {Bockstiegel}, \& {Walter}}]{Szypryt2016}
{Szypryt}, P., {Mazin}, B.~A., {Ulbricht}, G., {et~al.} 2016,
  \href{http://dx.doi.org/10.1063/1.4964665}{\JournalTitle{Applied Physics
  Letters}, 109, 151102}

\bibitem[{{Szypryt} {et~al.}(2014){Szypryt}, {Duggan}, {Mazin}, {Meeker},
  {Strader}, {van Eyken}, {Marsden}, {O'Brien}, {Walter}, {Ulbricht}, {Prince},
  {Stoughton}, \& {Bumble}}]{Szypryt2014}
{Szypryt}, P., {Duggan}, G.~E., {Mazin}, B.~A., {et~al.} 2014,
  \href{http://dx.doi.org/10.1093/mnras/stu137}{\JournalTitle{MNRAS}, 439,
  2765}

\bibitem[{Szypryt {et~al.}(2017)Szypryt, Meeker, Coiffard, Fruitwala, Bumble,
  Ulbricht, Walter, Daal, Bockstiegel, Collura, Zobrist, Lipartito, \&
  Mazin}]{Szypryt2017}
Szypryt, P., Meeker, S.~R., Coiffard, G., {et~al.} 2017,
  \href{http://dx.doi.org/10.1364/OE.25.025894}{\JournalTitle{Optics Express},
  25, 25894}

\bibitem[{{van Eyken} {et~al.}(2015){van Eyken}, {Strader}, {Walter}, {Meeker},
  {Szypryt}, {Stoughton}, {O'Brien}, {Marsden}, {Rice}, {Lin}, \&
  {Mazin}}]{vanEyken2015}
{van Eyken}, J.~C., {Strader}, M.~J., {Walter}, A.~B., {et~al.} 2015,
  \href{http://dx.doi.org/10.1088/0067-0049/219/1/14}{\JournalTitle{ApJS}, 219,
  14}

\bibitem[{Walter {et~al.}(2018)Walter, Bockstiegel, Mazin, \&
  Daal}]{Walter2017}
Walter, A.~B., Bockstiegel, C., Mazin, B.~A., \& Daal, M. 2018,
  \href{http://dx.doi.org/10.1109/TASC.2017.2773836}{\JournalTitle{IEEE
  Transactions on Applied Superconductivity}, 28, 1}

\bibitem[{{Zurlo} {et~al.}(2014){Zurlo}, {Vigan}, {Mesa}, {Gratton}, {Moutou},
  {Langlois}, {Claudi}, {Pueyo}, {Boccaletti}, {Baruffolo}, {Beuzit},
  {Costille}, {Desidera}, {Dohlen}, {Feldt}, {Fusco}, {Henning}, {Kasper},
  {Martinez}, {Moeller-Nilsson}, {Mouillet}, {Pavlov}, {Puget}, {Sauvage},
  {Turatto}, {Udry}, {Vakili}, {Waters}, \& {Wildi}}]{Zurlo2014}
{Zurlo}, A., {Vigan}, A., {Mesa}, D., {et~al.} 2014,
  \href{http://dx.doi.org/10.1051/0004-6361/201424204}{\JournalTitle{\aap},
  572, A85}

\end{thebibliography}


\end{document}